\documentclass[%
 aip,
 amsmath,amssymb,
 reprint,%
]{revtex4-1}

\usepackage{graphicx}
\usepackage{dcolumn}
\usepackage{bm}

\usepackage[utf8]{inputenc}
\usepackage[T1]{fontenc}
\usepackage{mathptmx}
\usepackage{mathtools}
\usepackage{etoolbox}
\usepackage{color}
\usepackage{hyperref}
\makeatletter
\def\@email#1#2{%
 \endgroup
 \patchcmd{\titleblock@produce}
  {\frontmatter@RRAPformat}
  {\frontmatter@RRAPformat{\produce@RRAP{*#1\href{mailto:#2}{#2}}}\frontmatter@RRAPformat}
  {}{}
}%
\makeatother
\begin{document}

\preprint{AIP/123-QED}

\title[]{Effective local potentials for density and density-matrix functional approximations with non-negative screening density}
\author{Thomas C. Pitts}
\affiliation{Department of Physics, Durham University, South Road, Durham, DH1 3LE, United Kingdom}
\author{Sofia Bousiadi}
 \affiliation{Theoretical and Physical Chemistry Institute, National Hellenic Research Foundation, Vass. Constantinou 48, GR-11635 Athens, Greece
}%
\affiliation{Faculty of Physics, National and Kapodistrian University of Athens, Panepistimiopolis, Zografos, Athens 157 84, Greece}
\author{Nikitas I. Gidopoulos}%
 \affiliation{Department of Physics, Durham University, South Road, Durham, DH1 3LE, United Kingdom}
\author{Nektarios N. Lathiotakis}
\email{lathiot@eie.gr}
\affiliation{Theoretical and Physical Chemistry Institute, National Hellenic Research Foundation, Vass. Constantinou 48, GR-11635 Athens, Greece
}%

\date{\today}

\begin{abstract}
A way to improve the accuracy of the spectral properties in density functional theory (DFT)
is to impose constraints on the effective, Kohn-Sham (KS), local potential [J. Chem. Phys. {\bf 136}, 224109 (2012)]. As illustrated, a convenient variational quantity in that approach is the ``screening'' or ``electron repulsion'' density, $\rho_{\rm rep}$, corresponding to the local, KS Hartree, exchange and correlation  potential through Poisson's equation. 
Two constraints, applied to this minimization, largely remove self-interaction errors from the effective potential: (i) $\rho_{\rm rep}$ integrates to $N-1$, where $N$ is the number of electrons, and (ii) $\rho_{\rm rep}\geq 0$ everywhere. 
In this work, we
introduce an effective ``screening'' amplitude, $f$, as the variational quantity, with the screening density being $\rho_{\rm rep}=f^2$. 
In this way, the positivity condition for $\rho_{\rm rep}$ is automatically satisfied and the minimization problem becomes more efficient and robust. 
We apply this technique to molecular calculations employing several approximations in DFT and in reduced density matrix functional theory. 
We find that the proposed development is an accurate, yet robust, variant of the constrained effective potential method. 
\end{abstract}

\maketitle

\section{\label{sec:level1}INTRODUCTION}
Replacing the Kohn-Sham (KS) potential \cite{KohnSham} for approximate functionals in Density Functional Theory 
(DFT) with a variational, constrained potential that fulfills exact properties is one of the proposed ways to improve spectral properties~\cite{gid} and more specifically the interpretation of KS orbital energies as a meaningful electronic spectrum of the system.  
One can minimize  the total energy of any Density Functional Approximation (DFA) under the constraint that the optimal effective local potential exhibits the correct asymptotic 
behavior. 
It was argued that, in this way, the effect of self-interactions 
(SI) on the local potential is suppressed and the ionization energies obtained by the orbital energies of the highest occupied molecular orbital (HOMO) 
are significantly closer to the experiment than those obtained from the unconstrained, approximate KS potential.~\cite{gid,tom,WOS:000370490900007,WOS:000603038300006}

In single particle theories, like KS, SIs can appear when the exact exchange energy functional is replaced with approximations that violate the complete cancellation of the Hartree SI terms. Characteristic examples are local/semi-local approximations like 
{LDA~\cite{KohnSham,per,CA1980,vwn} or 
GGAs.~\cite{GGA1,GGA2,GGA3,GGA4}} The seminal work of Perdew and Zunger~\cite{per} aimed at correcting this problem. Several other attempts were made to address the same problem.~\cite{lund,toh,goed1,gid,tom,van,legr,tsu,peders,gid2015,clark} All these attempts aim at correcting the total energy. Contrarily, the main concept of Ref.~\onlinecite{gid}, is to correct the effective potential instead through additional constraints. A typical 
manifestation of SIs is the wrong asymptotic behavior of the KS potential. On the other hand, the correct behavior of the potential at large distances plays a crucial role in the determination of several molecular properties. The Coulomb potential, $V_{\rm H}$, behaves like $N/r$ at large distances. In order to obtain the correct overall behavior, the exchange and correlation (xc) potential, $V_{\rm xc}$ should behave as $-1/r$, which does not hold for many approximations. Indeed, in most semilocal DFAs, the xc potential decays exponentially fast. 
In Ref.~[\onlinecite{gid}], the aim was to replace the Hartree-exchange and correlation (HXC) part of the KS potential with the energetically optimal effective potential $V_{\rm rep}$ (that minimizes the total energy functional) but which also satisfies two subsidiary conditions. These conditions or constraints are expressed in terms of the screening (or electron repulsion) 
density,~\cite{gorl} 
$\rho_{\rm rep}$; this density corresponds to 
$V_{\rm rep}$ through Poisson's equation. The first is that $\rho_{\rm rep}$ integrates to $N-1$ electrons and the second is that it is everywhere non-negative ($\rho_{\rm rep}({\bf r})\geq 0$). With this approach, the effect of SIs on the effective potential $V_{\rm rep}$ was largely corrected and molecular properties were improved.~\cite{gid,tom,WOS:000370490900007,WOS:000603038300006} 

A recent publication~\cite{WOS:000603038300006} explored the relaxation of the positivity constraint, $\rho_{\rm rep}({\bf r})\geq 0$; 
this leads to a mathematical optimization problem that is strictly not well-posed. 
Nevertheless, in practice, the mathematical problems became evident only with 
large auxiliary basis sets, or for systems with few electrons.
In addition, a density inversion method, yielding the corresponding KS potential that satisfies the correct asymptotic behavior was introduced recently.\cite{inv1,inv2} 
Finally, a hybrid scheme was proposed~\cite{hybrid}, where the  local part of the KS potential is optimized under the aforementioned subsidiary conditions, yielding single-particle energies in excellent agreement with  experimental ionization energies, even for the core electrons. 
The hybrid potential scheme in Ref.~\onlinecite{WOS:000603038300006} is applicable to any semilocal DFA. 

The idea of optimizing the energy functional with respect to a constrained local potential has also been applied to approximations within the framework of the reduced, density-matrix functional theory (RDMFT),~\cite{lathiot,lathiot2} leading to an approach that was called local RDMFT (LRDMFT). In RDMFT, functionals are usually explicitly expressed in terms of the natural orbitals and their corresponding occupation numbers, i.e. the eigenfunctions and eigenvalues of the one-body, reduced, density-matrix (1RDM). Although the foundations of RDMFT were laid a long time ago,~\cite{gilb, zumb, levy} 
{it has received significant attention in the last couple of decades.~\cite{goed1,goed2,bu,BB3,lathiotakis2009density,ML,piris,ext1,ext2,ext3}} One of its main advantages is that the kinetic energy can be written as an explicit functional of the 1-RDM, so the exchange-correlation term is not ``contaminated" by  kinetic energy contributions. Fractional occupation numbers are introduced through Coleman's $N$-representability conditions.~\cite{coleman} RDMFT provides the opportunity to express energy terms exactly as functionals of 1-RDM except for the electron-electron interaction energy. For this term, many approximate functionals have been proposed,~\cite{goed2,bu,fun4,lathiotakis2009density,fun5,fun6,pow,ML,BB0,pn5,BB3,PhysRevLett.127.233001} offering a good description of electronic correlations. In these approaches,
energy minimization is performed iteratively in two steps: (1) Minimization with respect to natural occupation numbers, and (2) minimization with respect to natural orbitals. The first of these steps is relatively inexpensive computationally, but the second is very demanding. Although significant progress has been achieved in improving its efficiency,~\cite{piris} this minimization still remains a considerable bottleneck. 

In LRDMFT,\cite{lathiot,lathiot2} functionals of the 1RDM are employed, but the minimization with respect to the orbitals is replaced with a constrained minimization that restricts the domain of orbital variation to those that are obtained from a single particle Schr\"odinger equation with local potential. Thus, it becomes a direct application of the constrained potential optimization of Refs.~\onlinecite{gid,tom,WOS:000370490900007,WOS:000603038300006} to 1RDM functionals.  
One important attribute of this approach is, of course, the relative efficiency in orbital optimization compared to full RDMFT minimization. A second one is that single-particle properties are offered through the spectrum of the single-particle Hamiltonian. It was found that energy eigenvalues of the highest occupied orbitals (HOMO) obtained in that way reproduce accurately the ionization potentials (IPs).\cite{lathiot,lathiot2} LRDMFT should not be considered as an approximation of RDMFT because the true natural orbitals cannot be obtained (even approximately) by a local potential and the optimal orbitals from LRDMFT are KS-like orbitals~\cite{theophilou2015orbitals}. It can be considered as a DFA with fractional occupation numbers introduced through the minimization of the energy. LRDMFT incorporates good aspects of both RDMFT and DFT, combining the low computational cost of DFT and the description of the static correlation of RDMFT through the introduction of fractional occupation numbers.

A hurdle in the application of the constrained minimization method is the enforcement of the positivity condition for the effective density, $\rho_{\rm rep}({\bf r})$. The fact that this condition has to be satisfied at each point in space, e.g. by a penalty term, has been proven computationally inefficient. In this work, we present a variant of the constrained minimization method with the positivity condition enforced by expressing the screening density as the square of an orbital-like ``screening-density amplitude'',
$\rho_{\rm rep}({\bf r})=f({\bf r})^2$, 
and using this amplitude, $f({\bf r})$, as the minimization variable. 
In that way, the positivity condition is automatically satisfied, however, there is a price to pay, since the variation with respect to $f({\bf r})$ does not lead to linear equations. Hence, in this work, we determine $f$ through a minimization procedure instead of solving any variational equations. 
In this first application, for simplicity, we assume that $f$ is expanded on the same orbital basis as the molecular orbitals. We apply the new method to the LDA functional in DFT as well as several approximations in LRDMFT. 

This paper is organized as follows: In Section~\ref{sec:theory}, we present the methodology to determine the screening-density amplitude. Section~\ref{sec:app} is devoted to applications. In 
{Section}~\ref{sec:app_dft}, we include the results of the application of the present method in optimizing total energies for a few local, semi-local, and hybrid DFAs and focus on the agreement of the obtained IPs with the experiment. Finally, in 
{Section}~\ref{sec:app_lrdmft}, we show the results of applying the present method to RDMFT functionals, i.e. as a variant of the LRDMFT methodology.

\section{\label{sec:theory}THEORY}
{In KS theory\cite{KohnSham}, the local potential $V_{\rm KS}$ is the sum $V_{\rm KS} = V_{\rm ext} +V_{\rm Hxc}$  of the external potential, $V_{\rm ext}$, and the Hartree, exchange, and correlation term, $V_{\rm Hxc}$. 
It is obtained from the functional derivative of the total electronic energy with respect to the electron density. This potential is inserted in the single particle
Kohn-Sham equations,
\begin{equation}\label{sch}
\Big[-\frac{\nabla^2}{2}+V_{\rm ext}({\bf r})+V_{\rm Hxc}({\bf r})\Big]\phi_{i}({\bf r} )=\epsilon_{i}\,\phi_{i}({\bf r} )\,,
\end{equation}
yielding the KS orbitals, hence the density, in a self-consistent cycle.}

{Following Refs.~\onlinecite{gid,tom}, $V_{\rm Hxc}$, in Eq.~(\ref{sch}), is replaced  by a 
variational effective repulsive potential $V_{\rm rep}$. 
The total energy, $E$, depending explicitly on the orbitals, $\{\phi_{i}\}$, can be considered a 
functional of $V_{\rm rep}$ if we assume that the orbitals are obtained from the solution of an equation like Eq.~(\ref{sch}), but with the variational potential, $V_{\rm rep}$, replacing $V_{\rm Hxc}$. For any DFA, this energy functional can
be minimized with respect to $V_{\rm rep}$, under additional constraints (subsidiary conditions), e.g. in order to enforce exact properties for the potential. Without any such constraint,
it is easy to see that the optimal $V_{\rm rep}$ will be equal to $V_{\rm Hxc}$. Since the considered additional constraints   concern the "screening" or "electron repulsion" density, $\rho_{\rm rep}$, associated with $V_{\rm rep}$ through Poisson's equation, 
it is convenient to express $V_{\rm rep}$ in terms of 
 $\rho_{\rm rep}$,
as~\cite{gorl,gid,tom,lathiot}
\begin{equation}\label{vrep}
    V_{\rm rep}({\bf r}) = \int d\bf{r}^\prime \, \frac{\rho_{\rm rep}(\bf{r}^\prime)}{|r-r^\prime|}\,.
\end{equation}
}

{An exact property of the KS potential is that the asymptotic behavior of the electron-electron repulsion part,
$V_{\rm Hxc}$, is}
\begin{equation}\label{eq:ab}
\lim_{r \rightarrow\infty} V_{\rm Hxc}({\bf r}) = \frac{N-\alpha}{r},\mbox{with\ } \alpha=1 \,,
\end{equation}
i.e. an electron at infinity ``feels" the repulsion of the remaining $N-1$ electrons.
One of the manifestations of SIs, which are present in most DFAs, is that this property is not satisfied by the approximate $V_{\rm Hxc}$, typically with
$0 \leq \alpha < 1$. 
In Ref.~\onlinecite{gid}, it was proposed that properties of the exact KS potential, like Eq.~(\ref{eq:ab}), can be enforced in the minimization with respect to $V_{\rm rep}$, thus the optimal $V_{\rm rep}$ differs from $V_{\rm Hxc}$, i.e. it is no longer the functional derivative of the total energy with respect to the density. For such an optimization, it is convenient to express $V_{\rm rep}$ in the  form\cite{gorl,gid} of Eq.~(\ref{vrep}).

 Assuming the form of Eq.~(\ref{vrep}) one can minimize the total DFA energy with respect to $\rho_{\rm rep}$, instead of the effective potential. $\rho_{\rm rep}$ can be expanded in an auxiliary basis set, in principle different than the basis used for the expansion of the KS orbitals. In terms of $\rho_{\rm rep}$, the condition (\ref{eq:ab}),
concerns the total repulsive charge, $Q_{\rm rep}$, 
\begin{equation}\label{N-1}
    Q_{\rm rep}=\int d{\bf r}\:\rho_{\rm rep}({\bf r})=N-1.
\end{equation}
In Refs.~\onlinecite{gid,tom}, it was argued that this condition alone does not lead to a mathematically well-posed optimisation problem and is not sufficient to obtain well-behaved solutions in general. 
The reason is that with this condition alone, the constrained 
minimization of the total energy of a DFA that is contaminated with SIs 
(i.e., whose unconstrained screening density corresponds to $Q_{\rm rep}^{\rm DFA}=N$) 
would not lead to well-converged solution. 
It would be energetically favorable for the constrained 
minimization to converge to a screening density 
that locally, near the system, matches the unconstrained screening density, 
integrating locally to screening charge $N$; 
the constraint $Q_{\rm rep}=N-1$ would then be satisfied by distributing negative screening 
charge equal to $-1$ away from the system, depending on the size of the auxiliary basis.
In Ref.~\onlinecite{WOS:000603038300006}, an investigation of the constrained minimization is shown, imposing the constraint just on the norm of the screening charge \eqref{N-1}.
 A proposed successful solution in Refs.~\onlinecite{gid,tom,WOS:000370490900007} is to supplement this condition with the positivity of $\rho_{\rm rep}$ at all points in space:
\begin{equation}\label{pos1}
    \rho_{\rm rep}({\bf r}) \geq 0, \quad \forall {\bf r}.
\end{equation}
This condition is sufficient to yield well-behaved 
solutions, although it is not necessarily satisfied by the exact KS potential. 
It concerns the whole Hxc screening  density, which is mostly positive because the Hartree repulsion dominates.
 So, according to this condition, the xc part of the screening density, can not 
exceed in absolute size the electronic density at any point in space. Unfortunately, this supplementary condition results in an additional computational cost since it has to 
be verified and enforced for all points in space. 

A choice that renders the positivity condition redundant, is to introduce a new variational
parameter, namely an effective screening-density amplitude, $f$, such that
\begin{equation}\label{pos2}
    \rho_{\rm rep}({\bf r})=f({\bf r})^2.
\end{equation}
\begin{table}
\caption{\label{table:DFA_pVTZ}
Ionization potentials, in eV, using cc-pVTZ basis sets, for various molecules obtained with different DFAs (a) with the standard KS scheme and (b) using the optimization of the screening-density amplitude proposed here, compared with vertical experimental (Exp) ionization potentials.~\cite{nist} Hartree-Fock (HF) Koopmans' results are also shown. For each DFA, the average absolute percentage error, $\Delta=({100}/{N})\sum_{i}|{(\chi_{i}-\chi_{i}^{\rm (Exp)})}/{\chi_{i}^{\rm (Exp)}}|$, is also included.} 
\centering
\begin{ruledtabular}
\begin{tabular}{c c c c c c c c c}
System&\multicolumn{2}{c}{LDA}&\multicolumn{2}{c}{PBE}&\multicolumn{2}{c}{B3LYP}& HF & Exp \\
 & (a) & (b) & (a) & (b) & (a) & (b) &&\\ \hline
 He & 15.30 & 23.13 & 15.63 & 23.65 & 17.91 & 23.77 & 24.97 & 24.59 \\
 Ne & 13.01 & 18.96 & 12.79 & 19.15 & 15.18 & 19.66 & 23.01 & 21.60 \\
 Be & 5.62 & 8.54 & 5.65 & 8.74 & 6.36 & 8.62 & 8.42 & 9.32 \\
 H\textsubscript{2} & 10.21 & 15.50 & 10.38 & 15.83 & 11.84 & 15.79 & 16.21 & 15.43 \\
 H\textsubscript{2}O  & 6.85 & 11.28 & 6.72 & 10.93 & 8.39 & 11.26 & 13.73 & 12.78 \\
 NH\textsubscript{3}& 5.9 & 9.82 & 5.84 & 9.52 & 7.23 & 9.82 & 11.64 & 10.80\\
CH\textsubscript{4} & 9.36 & 12.90 & 9.38 & 12.60 & 10.72 & 12.83 & 14.82 & 13.60 \\
O\textsubscript{3}& 8.03 & 10.86 & 7.83 & 11.00 & 9.60 & 11.46 & 13.18 & 12.73 \\
C\textsubscript{2}H\textsubscript{2} & 7.26 & 10.53 & 7.10 & 10.24 & 8.09 & 10.51 & 11.07 &11.49 \\
C\textsubscript{2}H\textsubscript{4} & 6.86 & 9.85 & 6.69 & 9.60 & 7.58 & 9.61 & 10.24 & 10.68\\
 C\textsubscript{2}H\textsubscript{5}& 5.75 & 8.61 & 5.69 & 8.31 & 6.92 & 8.67 & 10.73 & 9.85 \\
 SiH\textsubscript{4} & 8.47 & 11.28 & 8.51 & 11.39 & 9.68 & 11.35 & 13.23 & 11.00 \\
 H\textsubscript{2}O\textsubscript{2} & 6.10 & 9.78 & 5.97 & 9.31 & 7.68 &  9.82 & 13.08 & 11.70 \\ 
 O\textsubscript{2} & 5.95 & 9.63 & 5.70 & 9.63 & 7.04 & 9.95 & 12.78 & 12.30\\
 CO\textsubscript{2} &  9.19 & 12.37 & 8.97 & 12.25 & 10.35 & 12.41 & 14.74 &13.78 \\
 CO &  9.09 & 12.56 & 9.03 & 12.6 & 10.56 & 12.96 & 15.14 &14.01 \\
 Li\textsubscript{2} &  3.23 & 5.14 & 3.22 & 5.16 & 3.65 & 5.10 & 4.89 & 5.11 \\
 CH\textsubscript{3}OH & 6.07 & 9.42 & 5.99 & 9.13 & 7.50 & 9.31 & 12.22 & 10.96 \\
 C\textsubscript{2}H\textsubscript{6}& 8.01 & 11.01 & 8.07 & 10.80 & 9.36 & 11.3 & 13.22 & 11.99 \\
 CH\textsubscript{3}NH\textsubscript{2} & 5.24 & 8.43 & 5.20 & 7.99 & 6.52 & 8.27 & 10.66 & 9.65 \\
 C\textsubscript{2}H\textsubscript{5}OH & 6.03 & 8.85 & 5.95 & 8.39 & 7.43 & 8.84 & 11.99 & 10.00 \\
 \hline
 $\Delta$ &  38.87 & 9.74 & 39.4 & 11.21 & 28.21 & 9.14 & 8.31 & \\
\end{tabular}
\end{ruledtabular}
\end{table}
\begin{table}
\caption{\label{table:DFA_pVQZ}Same as in Table~\ref{table:DFA_pVTZ}, but for cc-pVQZ basis sets.}
\centering 
\begin{ruledtabular}
\begin{tabular}{c c c c c c c c c}
System&\multicolumn{2}{c}{LDA}&\multicolumn{2}{c}{PBE}&\multicolumn{2}{c}{B3LYP}& HF & Exp \\
 & (a) & (b) & (a) & (b) & (a) & (b) &&\\ \hline
 He & 15.32 & 23.13 & 15.66 & 23.59 & 17.93 & 23.73 & 24.98 & 24.59 \\
 Ne & 13.23 & 19.06 & 13.03 & 18.77 & 15.38 & 19.34 & 23.10 & 21.60 \\
 Be &  5.62 & 8.52 & 5.65 & 8.77 & 6.36 & 8.62 & 8.42 & 9.32 \\
 H\textsubscript{2} & 10.22 & 15.48 & 10.39 & 15.81 & 11.84 & 15.80 & 16.21 & 15.43 \\
 H\textsubscript{2}O &  7.07 & 11.33 & 6.94 & 11.11 & 8.57 & 11.49 & 13.82 &12.78 \\
 NH\textsubscript{3}&  6.06 & 9.88 & 5.99 & 9.66 & 7.36 & 9.96 & 11.69 & 10.80 \\
 CH\textsubscript{4}&  9.37 & 12.98 & 9.40 & 12.67 & 10.73 & 13.01 & 14.83 & 13.60 \\
 O\textsubscript{3}&  8.12 & 11.41 & 7.93 & 11.36 & 9.68 & 11.73 & 13.22 & 12.73 \\
C\textsubscript{2}H\textsubscript{2}& 7.29 & 10.54 & 7.14 &10.13 & 8.12 & 10.52 & 11.10 & 11.49 \\
CO\textsubscript{2} &  9.24 & 12.86 & 9.03 & 12.23 & 10.41 & 12.84 & 14.79 & 13.78\\
Li\textsubscript{2}& 3.23 & 5.13 & 3.22 & 5.17 & 3.65 & 5.12 & 4.90 & 5.11\\
N\textsubscript{2}& 10.34 & 14.10 & 10.22 & 13.77 & 11.94 & 14.16 & 16.33 & 15.58 \\ \hline
$\Delta$& 37.15 & 7.19 & 37.54 & 8.55 & 27.21 & 6.54 & 6.02 & \\
\end{tabular}
\end{ruledtabular}
\end{table}

In order to get the 
derivative of the total energy, $E$, with respect to $f$,  we apply a chain rule and obtain
\begin{equation}\label{minim}
\frac{\delta E[f]}{\delta f({\bf x})} = 2 \,
f({\bf x})\left[ \Tilde{b}({\bf x})
-\int d{\bf y} \; \Tilde{\chi}({\bf x},{\bf y})\,f^2({\bf y})
\right]\,,  
\end{equation}
where 
\begin{equation}
\Tilde{b}({\bf x})= \int d{\bf r} \frac{1}{|{\bf r}-{\bf x}|} 
\int d{\bf r' } \;\chi({\bf r},{\bf r'}) \;V_{\rm Hxc}({\bf r' }) \,,   
\end{equation}
and

\begin{equation}
\tilde{\chi}({\bf x},{\bf y})=\iint d {\bf r} \,d{\bf r'} \frac{\chi({\bf r},{\bf r'})}{|{\bf r}-{\bf x}||{\bf r}'-{\bf y}|}\,.
\end{equation}
The quantity $\chi({\bf r},{\bf r'})$ is the KS density-density response function given by 
\begin{equation}
 \chi({\bf r},{\bf r}')=2\sum_{i}\sum_{\alpha}\frac{\phi_{i}({\bf r})\phi_{i}({\bf r})\phi_{\alpha}({\bf r}')\phi_{\alpha}({\bf r}')}{\epsilon_{i}-\epsilon_{\alpha}}\,.   
\end{equation}
Due to its similarity with an orbital, we can expand the amplitude $f$ in the same local basis set employed to expand the molecular orbitals, $f({\bf r})= \sum_{k} f_{k}\,\xi_{k} ({\bf r}) $ and we obtain:
\begin{equation}\label{matrix}
\frac{\partial E}{\partial f_n}= 2\sum_{k}B_{nk}\,f_{k} - 
2\sum_{kml}f_k \, A_{nkml} \,f_{m}\,f_{l}\,,
\end{equation}
where,
\begin{equation}
   A_{nkml}= \iint d{\bf x} \,d{\bf y}  \,\xi_{n}({\bf x})\,\xi_{k}({\bf x})\, 
  \tilde{\chi}({\bf x},{\bf y}) \,\xi_{m}({\bf y})\,\xi_{l}({\bf y}) \,,
\end{equation}
\begin{equation}
B_{nk}=\int d{\bf x} \;\xi_{n}({\bf x}) \,\Tilde{b}({\bf x}) \,\xi_{k}({\bf x}) \, .
\end{equation}

Eq.~(\ref{matrix}) no longer leads to a linear
equation unlike the equation derived in our previous work~\cite{gid}. Here, we solve directly the minimization problem of $E$ using standard minimization techniques\cite{gsl} which require the gradient with respect to $f_n$ of Eq.~(\ref{matrix}).

The optimization with respect to $f$ has to be carried out in the domain of amplitudes normalized to $N-1$ according to the constraint of Eq.~(\ref{N-1}). This requirement can be incorporated by 
{Lagrange multipliers technique. However, in the present work, we chose an alternative equivalent way by }
introducing an auxiliary amplitude, $f^\prime({\bf r})$, as our variational quantity that is not normalized, i.e.,
\begin{equation}\label{eqfprime}
    f_{n}=s\;f_{n}^{\prime}\,, \quad s=\frac{\sqrt{N-1}}{\left\Vert f^\prime \right\Vert}\,,
\end{equation}
where $\left\Vert f^\prime \right\Vert^2= \int d{\bf x} \, 
\big( f'({\bf x}) \big)^2 = 
\sum_{kl} f_{k}^\prime \, S_{kl}\, f_{l}^\prime$ and
$S_{kl}=\int d{\bf r}\;\xi_k({\bf r})\,\xi_l({\bf r})$.
{Then the gradient with respect to $f^\prime$ is
\begin{equation}\label{eq.grad.f'}
\frac{\partial E}{\partial f_{n}^\prime}=  s \frac{\partial E}{\partial f_{n}} 
-\frac{s}{\left\Vert f^\prime\right\Vert^2}
\sum_{m}\frac{\partial E}{\partial f_{m}}
f_m^\prime \sum_{\lambda} S_{n \lambda} f_{\lambda}^\prime\,.
\end{equation}
Through Eq.~(\ref{eqfprime}), $E$ become as functional of $f'$,  $E=E[f]=E[s\,f^\prime]$, and can be minimized through an unconstrained variation with respect to $f^\prime$, using the gradient with respect to $f^\prime$ given in Eq.~(\ref{eq.grad.f'}). 
}

In principle, one can expand the auxiliary amplitude $f^\prime$ (or equivalently $f$) in any set of basis functions, different from that of the orbitals. In this first implementation, however, we chose to expand it in the same basis set as the orbitals. This is a restriction compared to our previous work, where $\rho_{\rm rep}$ was expanded in a different (auxiliary) basis set. However, due to the similarity of the screening-density amplitude to an orbital, as we will demonstrate, this choice does not affect the accuracy of our results. We implemented the present method in the HIPPO computer code.~\cite{hip} 
\section{Applications\label{sec:app}}
\subsection{Application to DFAs\label{sec:app_dft}}
As a first application, we use the effective screening density amplitude method for the constrained optimization of DFAs for molecular systems. The effect of applying this method to common DFAs is to improve the values of the ionization potentials as calculated from the highest occupied molecular orbital (HOMO) energies in the same manner as in Refs.~\onlinecite{gid,tom}.

Our results for several molecules and different DFAs are presented in Tables \ref{table:DFA_pVTZ} and \ref{table:DFA_pVQZ} using the cc-pVTZ and cc-pVQZ basis sets, respectively, compared with results from plain DFAs and 
{the experiment values of (vertical) IPs obtained from the NIST WebBook~\cite{nist}}. As we see, the corrected DFAs have an error of roughly half that of the uncorrected functionals. These corrected IPs have a similar error to those from Hartree-Fock without needing to perform a non-local calculation. As mentioned the screening-density amplitude is expanded on the same basis as the orbitals and that introduces an additional basis set dependence of the IPs. Reasonably, the results of the larger, cc-pVQZ basis sets, are superior.

\begin{figure}[ht]
    \centering    \includegraphics[width=0.98\columnwidth]{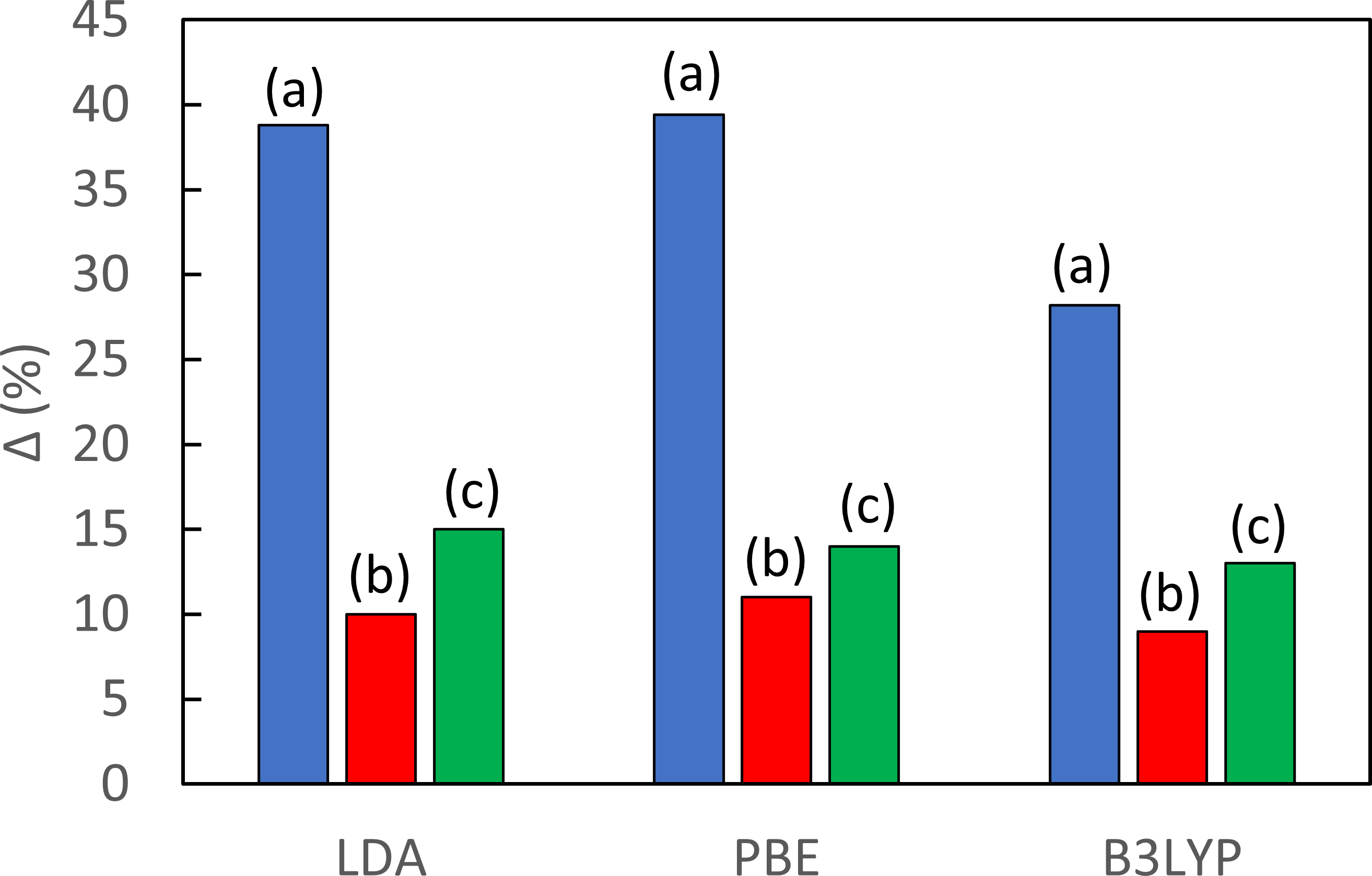}
      \caption{\label{comp_del}
      Average absolute percentage error, $\Delta$, defined in the caption of Table~\ref{table:DFA_pVTZ}, in the IPs for various DFAs calculated with (a) the standard KS scheme, (b) the present method, and (c) results from Refs.~\onlinecite{gid,tom}.}
\end{figure}

The present results are in line with those of our previous work \cite{gid,tom} where the $\rho_{\rm rep}$ was the variational parameter and the positivity had to be enforced additionally. This is shown in Fig.~\ref{comp_del} where we compare the average absolute percentage errors for the present method with those of Ref.~\onlinecite{tom}. However, we should note that the results of Ref.~\onlinecite{tom} are obtained with cc-pVDZ orbital basis and an auxiliary uncontracted cc-pVDZ basis for the screening density. The  present method has enhanced convergence compared to that of Ref~\onlinecite{gid} in the objective functional minimization and provides a more robust framework. 


\subsection{Application to Local-RDMFT\label{sec:app_lrdmft}}
As a second application of the present constrained minimization we considered LRDMFT, i.e. we minimized functionals of the  1RDM under the subsidiary condition that the orbitals are eigenfunctions of a single particle Hamiltonian with local potential.  The repulsion part of this potential is assumed to be of the form of Eq.~(\ref{vrep}) where $\rho_{\rm rep}$ is given by the quadratic form 
of Eq.~(\ref{pos2}). The occupation numbers are minimized directly as in standard RDMFT calculations.

For the calculations, we employed five popular approximate RDMFT functionals namely Mueller,~\cite{BB0} BBC3,~\cite{BB3} Power,~\cite{pow} ML,~\cite{ML} and PNOF5.~\cite{pn5} 
As far as the basis sets are concerned the cc-pVTZ and cc-pVQZ were employed, for expanding both the orbitals and the screening-density amplitude, for selected systems. 
{However, due to limitations of the HIPPO code, g-functions were removed from the cc-pVQZ basis set, a modification that is not expected to affect our results. }
 
\begin{figure}[ht]
    \centering
    \includegraphics[width=0.95\columnwidth]{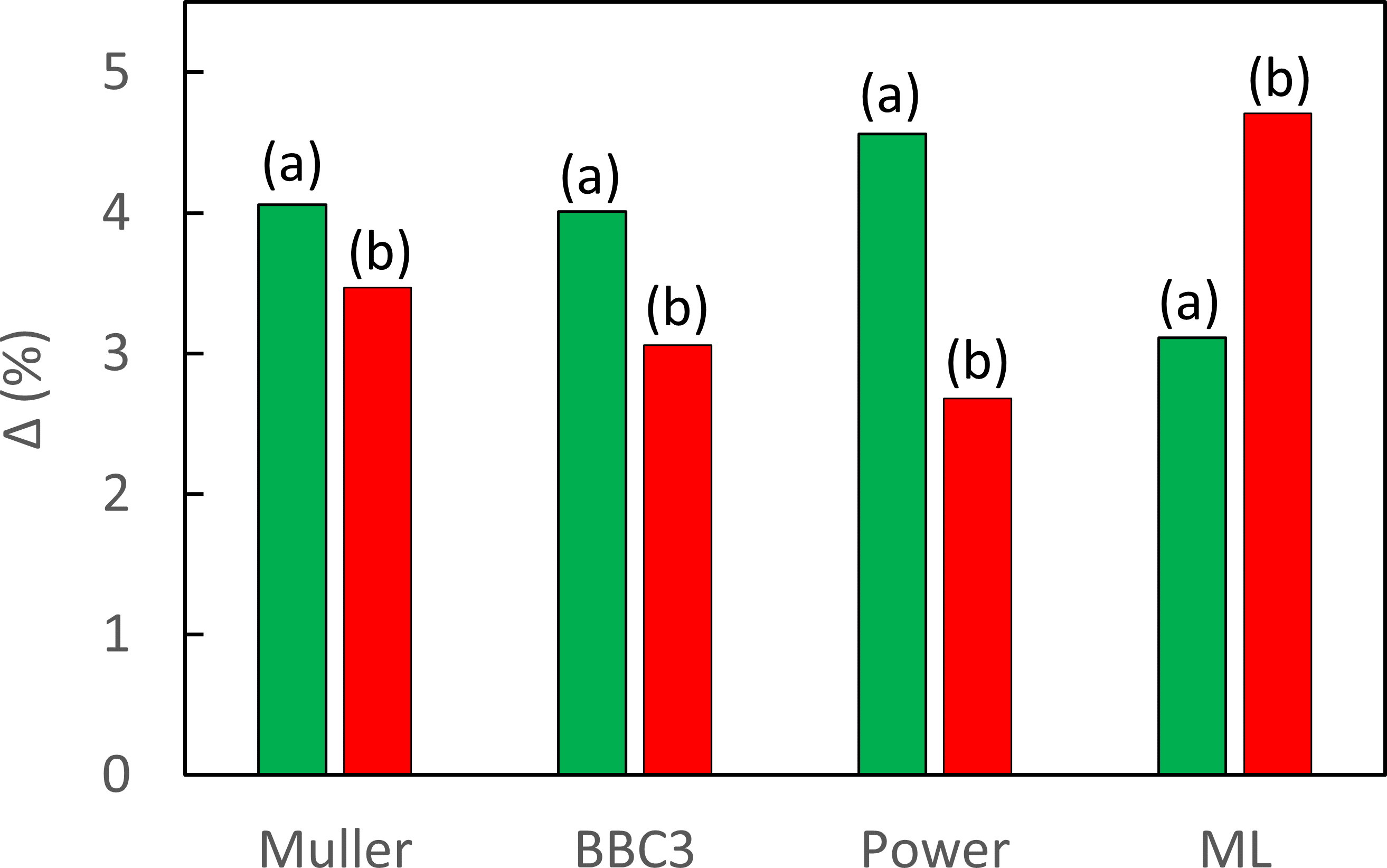}
      \caption{\label{fig:rdmft}
      Average absolute percentage error, $\Delta$ defined in the caption of Table~\ref{table:DFA_pVTZ}, in the IPs calculated with LRDMFT for various functionals with (a) the method of Ref. \onlinecite{lathiot} and (b) the present method.}
\end{figure}

We compare energy eigenvalues of the HOMOs with the values of (vertical) experimental IPs 
{(from NIST WebBook \cite{nist}), in Tables~\ref{rd1} and \ref{rd2} for the two different basis sets, respectively}.  For comparison, we also calculated IPs from standard Hartree-Fock through Koopman's theorem. As we see, there is a general trend of improvement of the results with the size of the basis set, i.e. the results of the cc-PVQZ basis set are better than those of cc-pVTZ. Nevertheless, our results are of similar quality to those of Refs.~\onlinecite{lathiot,lathiot2}, as seen in Fig.~\ref{fig:rdmft} and Table~\ref{rd3}. As we see, BBC3, Power and PNOF5 functionals are yielding very good results compared with the experiment. 
\begin{table}
\caption{\label{rd1}
Calculated ionization potentials, in eV, for cc-pVTZ basis sets, for various molecules, using the present variant of LRDMFT of optimizing the screening-density amplitude and different functionals of the 1RDM, compared with vertical experimental ionization potentials~\cite{nist}. Hartree-Fock Koopmans' results are also shown. The average absolute percentage errors, defined in the caption of Table~\ref{table:DFA_pVTZ}, are also included.}
\centering
\begin{ruledtabular}
\begin{tabular}{c c c c c c c c}
System &  Müller & BBC3 & POW & ML & PNOF5 & HF & Exp.  \\[0.5ex]
\hline
He & 24.67 & 24.62 & 24.78 & 25.05 & 24.35 & 24.97 & 24.59 \\
Ne & 21.76 & 22.34 & 22.22 & 22.76 & 22.58 & 23.01 & 21.6 \\
Be & 9.41 & 8.93 & 8.80 & 8.74 & 8.72 & 8.42 & 9.32  \\
H\textsubscript{2} & 16.27 & 16.18 & 16.16 & 16.23 &16.14 & 16.21 & 15.43 \\
H\textsubscript{2}O & 13.19 & 12.9 & 12.89 & 13.39 &13.39 & 13.73 &12.78 \\
NH\textsubscript{3} & 11.62 & 11.12 & 11.13 &11.36 & 11.31 & 11.64 & 10.80 \\
CH\textsubscript{4} & 14.11 & 14.10 & 13.91 & 14.23 & 14.12 & 14.82 & 13.60 \\
O\textsubscript{3} & 14.02 & 12.87 & 14.10 & 13.19 & 12.42 & 13.18 & 12.73 \\
C\textsubscript{2}H\textsubscript{2} &  12.20 & 11.26 & 11.47 & 11.62 & 11.50 & 11.07 & 11.49  \\
C\textsubscript{2}H\textsubscript{4} & 10.98 & 10.59 & 10.73 & 10.92 & 10.74 & 10.24 & 10.68 \\
C\textsubscript{2}H\textsubscript{5}N & 11.33 & 10.40 & 10.53  & 10.37 & 10.37 & 10.73 & 9.85 \\
SiH\textsubscript{4} & 11.33 & 11.72 & 10.39  & 10.38 & 11.07 & 13.24 &11.00  \\
H\textsubscript{2}O\textsubscript{2} & 11.31 & 11.70 & 11.18 & 12.25 & 11.53 &13.08 & 11.70 \\
O\textsubscript{2} & 13.04 & 11.9 & 12.98 & 11.90 & 12.24 &12.78 & 12.30 \\
CO\textsubscript{2} & 14.29 & 14.82 & 14.61 & 15.23 & 14.86 & 14.74 & 13.78 \\
CO & 14.44 & 14.57 & 14.79  & 13.73 & 14.37 & 15.14 &14.01 \\
Li\textsubscript{2} & 5.96 & 4.99 & 5.06 & 4.91 & 4.87 & 4.89 & 5.11  \\
CH\textsubscript{3}OH & 11.00 & 10.93 & 11.00 & 11.20 & 11.07 & 12.22 &10.96  \\
C\textsubscript{2}H\textsubscript{6} & 12.01 & 12.29 & 12.11 & 12.53 & 12.24 &13.22 & 11.99  \\
CH\textsubscript{3}NH\textsubscript{2} & 10.24 & 9.69 & 9.88 & 10.00 & 9.81 & 10.66 & 9.65 \\ 
C\textsubscript{2}H\textsubscript{5}OH & 10.35 & 9.78 & 10.80 & 10.61 & 10.00 & 11.97 & 10.00 \\ \hline
$\Delta$  & 5.04 & 2.98 & 4.17 &	4.70 & 3.30 & 8.5 \\
\end {tabular}
\end{ruledtabular}
\end{table}
\begin{table}
\caption{\label{rd2} Same as in Table~\ref{rd1}, for cc-pVQZ basis sets.
}
\centering 
\begin{ruledtabular}
\begin{tabular}{c c c c c c c}
System &  Müller & BBC3 & POW & PNOF5 & HF & Exp  \\[0.5ex]
\hline
He & 24.67 & 24.61 & 24.78 & 24.41 & 24.98 &24.59 \\
Ne & 21.59 & 22.3 & 22.24 & 22.72 & 23.10 & 21.60 \\
Be & 9.24 & 8.80 & 8.59 & 8.49 & 8.42 & 9.32  \\
H\textsubscript{2} & 16.13 & 16.11 & 16.11 & 16.13 & 16.21 & 15.43 \\
H\textsubscript{2}O &12.90 & 13.15 & 12.60 & 13.09 &13.82 &12.78 \\
NH\textsubscript{3} & 11.60 & 11.15 & 10.97 & 11.60 &11.69 &10.80 \\
CH\textsubscript{4} & 13.68 & 13.92 & 13.83 & 14.14 &14.83 &13.60 \\
O\textsubscript{3} & 13.54 & 12.42 & 13.58 & 12.53 &13.22 &12.73 \\
C\textsubscript{2}H\textsubscript{2} & 11.94 & 10.72 & 10.73 & 10.43 &11.10 &10.68 \\
CO\textsubscript{2} & 14.15 & 14.47 & 14.13 & 13.90 &14.79 &13.78 \\
Li\textsubscript{2} & 5.98 & 5.00 & 5.04 & 4.88 &4.90 &5.11  \\
N\textsubscript{2} & 15.91 & 15.99 & 16.36 & 15.60 &16.33 &15.58 \\
\hline
$\Delta$ 	& 4.56 &	2.87 &	3.06 & 3.55 &  6.10\\
\end {tabular}
\end{ruledtabular}
\end{table}
\begin{table*}
 \caption{\label{rd3}
 Ionization potentials, in eV, calculated with LRDMFT,  for various 1-RDM functionals, and cc-pVTZ basis sets, using (a) the method of Ref. \onlinecite{lathiot} and (b) the present method.}
 \begin{ruledtabular}
\begin{tabular}{lllllllllll}
     & Müller (a) & Müller (b) & BBC3 (a)  & BBC3 (b) & Power (a) &  Power (b) & ML (a)    & ML (b)& Exp.   &  \\
     \hline 
He   & 24.69  & 24.67  & 24.57 & 24.62  & 24.84 & 24.78  & 25.15 & 25.05  & 24.59 &  \\
Be   & 9.51   & 9.41  & 8.73  & 8.93 & 8.58  &  8.80 & 8.55  & 8.74 & 9.32 &  \\
Ne   & 22.9   & 21.76 & 20.92 & 22.34 & 21.65 & 22.22 & 21.32 & 22.76 & 21.6 &  \\
H\textsubscript{2}   & 16.24  & 16.27  & 16.15 & 16.18  & 16.15 & 16.16  & 16.28 & 16.23  & 15.43 &  \\
H\textsubscript{2}O  & 12.59  & 13.19  & 12.03 & 12.9   & 12.1  & 12.89  & 12.64 & 13.39  & 12.78 &  \\
NH\textsubscript{3}  & 11.03  & 11.62  & 10.65 & 11.12  & 10.74 & 11.13  & 10.95 & 11.36  & 10.8  &  \\
CH\textsubscript{4}  & 13.55  & 14.11  & 13.72 & 14.1   & 13.43 & 13.91  & 13.84 & 14.23  & 13.6  &  \\
C\textsubscript{2}H\textsubscript{2} & 11.67  & 12.2   & 11.12 & 11.26  & 11.46 & 11.47  & 11.59 & 11.62  & 11.49 &  \\
C\textsubscript{2}H\textsubscript{4} & 10.68  & 10.98  & 10.45 & 10.59  & 10.47 & 10.73  & 10.9  & 10.92  & 10.68 &  \\
CO\textsubscript{2}  & 13.81  & 14.29  & 13.67 & 14.82  & 13.3  & 14.61  & 14.42 & 15.23  & 13.78 & \\
\end{tabular}
\end{ruledtabular}
\end{table*}

 \section{conclusion\label{sec:conc}}
In conclusion, we present a variant of the constrained local potential optimization method,~\cite{gid} where an effective screening-density amplitude is introduced, replacing the screening density as the variational parameter. With this choice, the positivity condition of the screening density is automatically satisfied, and thus, it does not need to be imposed. This condition was proven necessary in many cases to obtain reasonable effective local potentials without certain pathologies in the asymptotic region. However, its explicit imposition is a formidable task and in this work, we circumvent this problem by assuming a quadratic form for the screening density. The price to pay is that the optimization with respect to the effective amplitude is no longer equivalent to the solution of a set of linear equations. Although linearization schemes can be applied to solve these equations, we chose, in this first application, to solve the problem by a direct minimization algorithm. In addition, for simplicity, we expanded the effective amplitude in the same basis set as the orbitals. 

We demonstrate that, with the present method, accurate local potentials are obtained, as evidenced by the agreement of the obtained IPs for several DFAs with those of our previous work  and the experiment. Thus the choice of expanding the effective amplitude in the same basis set as the orbitals is not affecting the method's accuracy. This accuracy, as expected, improves with the size of the basis set. In addition, similarly, we applied
the effective amplitude method to local RDMFT, i.e. we minimized functionals of the 1RDM in terms of the effective amplitude. We found that the obtained IPs are of similar quality to those of Refs~\onlinecite{lathiot,lathiot2}, i.e. the present minimization technique offers a more efficient and robust method in LRDMFT.

{The pursuit of high-quality spectral properties coming from single particle Kohn-Sham theoretical models is of central importance and there exist numerous efforts addressing this issue.~\cite{HAMEL2002345,doi:10.1063/1.1430255,WOS:000343196300027,doi:10.1063/1.3678180,doi:10.1063/1.5116338,doi:10.1063/1.5084728} The methodology of applying additional constraints to the potential and in that way enforcing exact properties is an elegant and promising solution that is applicable to any DFA by modifying the potential but not the energy functional.~\cite{gid,tom,WOS:000370490900007,WOS:000603038300006} As such, it is fairly easy to implement in existing computer codes. If tuned we obtain very accurate ionization potentials compared to other theoretical models and experiment.~\cite{hybrid} With the present proposition, this methodology becomes simpler, more robust, and more efficient than previous implementations yet equally accurate. }

\begin{acknowledgments}
We acknowledge funding for this work by the Hellenic Foundation for Research
and Innovation (HFRI) under the HFRI Ph.D. Fellowship grant (Fellowship Number: 1310) and by 
The Leverhulme Trust, through a Research Project Grant with number RPG-2016-005.
\end{acknowledgments}

\nocite{*}
\bibliography{Effective_orb}

\begin{thebibliography}{57}%
\makeatletter
\providecommand \@ifxundefined [1]{%
 \@ifx{#1\undefined}
}%
\providecommand \@ifnum [1]{%
 \ifnum #1\expandafter \@firstoftwo
 \else \expandafter \@secondoftwo
 \fi
}%
\providecommand \@ifx [1]{%
 \ifx #1\expandafter \@firstoftwo
 \else \expandafter \@secondoftwo
 \fi
}%
\providecommand \natexlab [1]{#1}%
\providecommand \enquote  [1]{``#1''}%
\providecommand \bibnamefont  [1]{#1}%
\providecommand \bibfnamefont [1]{#1}%
\providecommand \citenamefont [1]{#1}%
\providecommand \href@noop [0]{\@secondoftwo}%
\providecommand \href [0]{\begingroup \@sanitize@url \@href}%
\providecommand \@href[1]{\@@startlink{#1}\@@href}%
\providecommand \@@href[1]{\endgroup#1\@@endlink}%
\providecommand \@sanitize@url [0]{\catcode `\\12\catcode `\$12\catcode
  `\&12\catcode `\#12\catcode `\^12\catcode `\_12\catcode `\%12\relax}%
\providecommand \@@startlink[1]{}%
\providecommand \@@endlink[0]{}%
\providecommand \url  [0]{\begingroup\@sanitize@url \@url }%
\providecommand \@url [1]{\endgroup\@href {#1}{\urlprefix }}%
\providecommand \urlprefix  [0]{URL }%
\providecommand \Eprint [0]{\href }%
\providecommand \doibase [0]{http://dx.doi.org/}%
\providecommand \selectlanguage [0]{\@gobble}%
\providecommand \bibinfo  [0]{\@secondoftwo}%
\providecommand \bibfield  [0]{\@secondoftwo}%
\providecommand \translation [1]{[#1]}%
\providecommand \BibitemOpen [0]{}%
\providecommand \bibitemStop [0]{}%
\providecommand \bibitemNoStop [0]{.\EOS\space}%
\providecommand \EOS [0]{\spacefactor3000\relax}%
\providecommand \BibitemShut  [1]{\csname bibitem#1\endcsname}%
\let\auto@bib@innerbib\@empty
\bibitem [{\citenamefont {Kohn}\ and\ \citenamefont {Sham}(1965)}]{KohnSham}%
  \BibitemOpen
  \bibfield  {author} {\bibinfo {author} {\bibfnamefont {W.}~\bibnamefont
  {Kohn}}\ and\ \bibinfo {author} {\bibfnamefont {L.~J.}\ \bibnamefont
  {Sham}},\ }\bibfield  {title} {\enquote {\bibinfo {title} {Self-consistent
  equations including exchange and correlation effects},}\ }\href {\doibase
  10.1103/PhysRev.140.A1133} {\bibfield  {journal} {\bibinfo  {journal} {Phys.
  Rev.}\ }\textbf {\bibinfo {volume} {140}},\ \bibinfo {pages} {A1133--A1138}
  (\bibinfo {year} {1965})}\BibitemShut {NoStop}%
\bibitem [{\citenamefont {Gidopoulos}\ and\ \citenamefont
  {Lathiotakis}(2012)}]{gid}%
  \BibitemOpen
  \bibfield  {author} {\bibinfo {author} {\bibfnamefont {N.~I.}\ \bibnamefont
  {Gidopoulos}}\ and\ \bibinfo {author} {\bibfnamefont {N.~N.}\ \bibnamefont
  {Lathiotakis}},\ }\bibfield  {title} {\enquote {\bibinfo {title}
  {Constraining density functional approximations to yield self-interaction
  free potentials},}\ }\href {\doibase 10.1063/1.4728156} {\bibfield  {journal}
  {\bibinfo  {journal} {The Journal of Chemical Physics}\ }\textbf {\bibinfo
  {volume} {136}},\ \bibinfo {pages} {224109} (\bibinfo {year}
  {2012})}\BibitemShut {NoStop}%
\bibitem [{\citenamefont {Pitts}, \citenamefont {Gidopoulos},\ and\
  \citenamefont {Lathiotakis}(2018)}]{tom}%
  \BibitemOpen
  \bibfield  {author} {\bibinfo {author} {\bibfnamefont {T.}~\bibnamefont
  {Pitts}}, \bibinfo {author} {\bibfnamefont {N.~I.}\ \bibnamefont
  {Gidopoulos}}, \ and\ \bibinfo {author} {\bibfnamefont {N.~N.}\ \bibnamefont
  {Lathiotakis}},\ }\bibfield  {title} {\enquote {\bibinfo {title} {Performance
  of the constrained minimization of the total energy in density functional
  approximations: the electron repulsion density and potential},}\ }\href
  {\doibase 10.1140/epjb/e2018-90123-8} {\bibfield  {journal} {\bibinfo
  {journal} {The European Physical Journal B}\ }\textbf {\bibinfo {volume}
  {91}},\ \bibinfo {pages} {1--9} (\bibinfo {year} {2018})}\BibitemShut
  {NoStop}%
\bibitem [{\citenamefont {Gidopoulos}\ and\ \citenamefont
  {Lathiotakis}(2015{\natexlab{a}})}]{WOS:000370490900007}%
  \BibitemOpen
  \bibfield  {author} {\bibinfo {author} {\bibfnamefont {N.}~\bibnamefont
  {Gidopoulos}}\ and\ \bibinfo {author} {\bibfnamefont {N.~N.~N.}\ \bibnamefont
  {Lathiotakis}},\ }\bibfield  {title} {\enquote {\bibinfo {title} {Constrained
  local potentials for self-interaction correction},}\ }in\ \href {\doibase
  10.1016/bs.aamop.2015.06.003} {\emph {\bibinfo {booktitle} {Advances In
  Atomic Molecular and Optical Physics, Chap. 6, Constrained Local Potentials
  for Self-Interaction Correction}}},\ \bibinfo {series} {Advances In Atomic
  Molecular and Optical Physics}, Vol.~\bibinfo {volume} {64},\ \bibinfo
  {editor} {edited by\ \bibinfo {editor} {\bibfnamefont {E.}~\bibnamefont
  {Arimondo}}, \bibinfo {editor} {\bibfnamefont {C.}~\bibnamefont {Lin}}, \
  and\ \bibinfo {editor} {\bibfnamefont {S.}~\bibnamefont {Yelin}}}\ (\bibinfo
  {year} {2015})\ pp.\ \bibinfo {pages} {129--142}\BibitemShut {NoStop}%
\bibitem [{\citenamefont {Callow}\ \emph {et~al.}(2020)\citenamefont {Callow},
  \citenamefont {Pearce}, \citenamefont {Pitts}, \citenamefont {Lathiotakis},
  \citenamefont {Hodgson},\ and\ \citenamefont
  {Gidopoulos}}]{WOS:000603038300006}%
  \BibitemOpen
  \bibfield  {author} {\bibinfo {author} {\bibfnamefont {T.~J.}\ \bibnamefont
  {Callow}}, \bibinfo {author} {\bibfnamefont {B.~J.}\ \bibnamefont {Pearce}},
  \bibinfo {author} {\bibfnamefont {T.}~\bibnamefont {Pitts}}, \bibinfo
  {author} {\bibfnamefont {N.~N.}\ \bibnamefont {Lathiotakis}}, \bibinfo
  {author} {\bibfnamefont {M.~J.~P.}\ \bibnamefont {Hodgson}}, \ and\ \bibinfo
  {author} {\bibfnamefont {N.~I.}\ \bibnamefont {Gidopoulos}},\ }\bibfield
  {title} {\enquote {\bibinfo {title} {Improving the exchange and correlation
  potential in density-functional approximations through constraints},}\ }\href
  {\doibase 10.1039/d0fd00069h} {\bibfield  {journal} {\bibinfo  {journal}
  {Faraday Discussions}\ }\textbf {\bibinfo {volume} {224}},\ \bibinfo {pages}
  {126--144} (\bibinfo {year} {2020})}\BibitemShut {NoStop}%
\bibitem [{\citenamefont {Perdew}\ and\ \citenamefont {Zunger}(1981)}]{per}%
  \BibitemOpen
  \bibfield  {author} {\bibinfo {author} {\bibfnamefont {J.~P.}\ \bibnamefont
  {Perdew}}\ and\ \bibinfo {author} {\bibfnamefont {A.}~\bibnamefont
  {Zunger}},\ }\bibfield  {title} {\enquote {\bibinfo {title} {Self-interaction
  correction to density-functional approximations for many-electron systems},}\
  }\href {\doibase 10.1103/PhysRevB.23.5048} {\bibfield  {journal} {\bibinfo
  {journal} {Physical Review B}\ }\textbf {\bibinfo {volume} {23}},\ \bibinfo
  {pages} {5048} (\bibinfo {year} {1981})}\BibitemShut {NoStop}%
\bibitem [{\citenamefont {Ceperley}\ and\ \citenamefont
  {Alder}(1980)}]{CA1980}%
  \BibitemOpen
  \bibfield  {author} {\bibinfo {author} {\bibfnamefont {D.~M.}\ \bibnamefont
  {Ceperley}}\ and\ \bibinfo {author} {\bibfnamefont {B.~J.}\ \bibnamefont
  {Alder}},\ }\bibfield  {title} {\enquote {\bibinfo {title} {Ground-state of
  the electron-gas by a stochastic method},}\ }\href {\doibase
  10.1103/PhysRevLett.45.566} {\bibfield  {journal} {\bibinfo  {journal}
  {Physical Review Letters}\ }\textbf {\bibinfo {volume} {45}},\ \bibinfo
  {pages} {566--569} (\bibinfo {year} {1980})}\BibitemShut {NoStop}%
\bibitem [{\citenamefont {Vosko}, \citenamefont {Wilk},\ and\ \citenamefont
  {Nusair}(1980)}]{vwn}%
  \BibitemOpen
  \bibfield  {author} {\bibinfo {author} {\bibfnamefont {S.~H.}\ \bibnamefont
  {Vosko}}, \bibinfo {author} {\bibfnamefont {L.}~\bibnamefont {Wilk}}, \ and\
  \bibinfo {author} {\bibfnamefont {M.}~\bibnamefont {Nusair}},\ }\bibfield
  {title} {\enquote {\bibinfo {title} {Accurate spin-dependent electron liquid
  correlation energies for local spin-density calculations - a critical
  analysis},}\ }\href {\doibase 10.1139/p80-159} {\bibfield  {journal}
  {\bibinfo  {journal} {Canadian Journal of Physics}\ }\textbf {\bibinfo
  {volume} {58}},\ \bibinfo {pages} {1200--1211} (\bibinfo {year}
  {1980})}\BibitemShut {NoStop}%
\bibitem [{\citenamefont {Perdew}\ \emph {et~al.}(1992)\citenamefont {Perdew},
  \citenamefont {Chevary}, \citenamefont {Vosko}, \citenamefont {Jackson},
  \citenamefont {Pederson}, \citenamefont {Singh},\ and\ \citenamefont
  {Fiolhais}}]{GGA1}%
  \BibitemOpen
  \bibfield  {author} {\bibinfo {author} {\bibfnamefont {J.~P.}\ \bibnamefont
  {Perdew}}, \bibinfo {author} {\bibfnamefont {J.~A.}\ \bibnamefont {Chevary}},
  \bibinfo {author} {\bibfnamefont {S.~H.}\ \bibnamefont {Vosko}}, \bibinfo
  {author} {\bibfnamefont {K.~A.}\ \bibnamefont {Jackson}}, \bibinfo {author}
  {\bibfnamefont {M.~R.}\ \bibnamefont {Pederson}}, \bibinfo {author}
  {\bibfnamefont {D.~J.}\ \bibnamefont {Singh}}, \ and\ \bibinfo {author}
  {\bibfnamefont {C.}~\bibnamefont {Fiolhais}},\ }\bibfield  {title} {\enquote
  {\bibinfo {title} {Atoms, molecules, solids, and surfaces: Applications of
  the generalized gradient approximation for exchange and correlation},}\
  }\href {\doibase 10.1103/PhysRevB.46.6671} {\bibfield  {journal} {\bibinfo
  {journal} {Phys. Rev. B}\ }\textbf {\bibinfo {volume} {46}},\ \bibinfo
  {pages} {6671--6687} (\bibinfo {year} {1992})}\BibitemShut {NoStop}%
\bibitem [{\citenamefont {Becke}(1988)}]{GGA2}%
  \BibitemOpen
  \bibfield  {author} {\bibinfo {author} {\bibfnamefont {A.~D.}\ \bibnamefont
  {Becke}},\ }\bibfield  {title} {\enquote {\bibinfo {title}
  {Density-functional exchange-energy approximation with correct asymptotic
  behavior},}\ }\href {\doibase 10.1103/PhysRevA.38.3098} {\bibfield  {journal}
  {\bibinfo  {journal} {Phys. Rev. A}\ }\textbf {\bibinfo {volume} {38}},\
  \bibinfo {pages} {3098--3100} (\bibinfo {year} {1988})}\BibitemShut {NoStop}%
\bibitem [{\citenamefont {Lee}, \citenamefont {Yang},\ and\ \citenamefont
  {Parr}(1988)}]{GGA3}%
  \BibitemOpen
  \bibfield  {author} {\bibinfo {author} {\bibfnamefont {C.}~\bibnamefont
  {Lee}}, \bibinfo {author} {\bibfnamefont {W.}~\bibnamefont {Yang}}, \ and\
  \bibinfo {author} {\bibfnamefont {R.~G.}\ \bibnamefont {Parr}},\ }\bibfield
  {title} {\enquote {\bibinfo {title} {Development of the colle-salvetti
  correlation-energy formula into a functional of the electron density},}\
  }\href {\doibase 10.1103/PhysRevB.37.785} {\bibfield  {journal} {\bibinfo
  {journal} {Phys. Rev. B}\ }\textbf {\bibinfo {volume} {37}},\ \bibinfo
  {pages} {785--789} (\bibinfo {year} {1988})}\BibitemShut {NoStop}%
\bibitem [{\citenamefont {Perdew}, \citenamefont {Burke},\ and\ \citenamefont
  {Ernzerhof}(1996)}]{GGA4}%
  \BibitemOpen
  \bibfield  {author} {\bibinfo {author} {\bibfnamefont {J.~P.}\ \bibnamefont
  {Perdew}}, \bibinfo {author} {\bibfnamefont {K.}~\bibnamefont {Burke}}, \
  and\ \bibinfo {author} {\bibfnamefont {M.}~\bibnamefont {Ernzerhof}},\
  }\bibfield  {title} {\enquote {\bibinfo {title} {Generalized gradient
  approximation made simple},}\ }\href {\doibase 10.1103/PhysRevLett.77.3865}
  {\bibfield  {journal} {\bibinfo  {journal} {Phys. Rev. Lett.}\ }\textbf
  {\bibinfo {volume} {77}},\ \bibinfo {pages} {3865--3868} (\bibinfo {year}
  {1996})}\BibitemShut {NoStop}%
\bibitem [{\citenamefont {Lundberg}\ and\ \citenamefont
  {Siegbahn}(2005)}]{lund}%
  \BibitemOpen
  \bibfield  {author} {\bibinfo {author} {\bibfnamefont {M.}~\bibnamefont
  {Lundberg}}\ and\ \bibinfo {author} {\bibfnamefont {P.~E.}\ \bibnamefont
  {Siegbahn}},\ }\bibfield  {title} {\enquote {\bibinfo {title} {Quantifying
  the effects of the self-interaction error in dft: When do the delocalized
  states appear?}}\ }\href {\doibase 10.1063/1.1926277} {\bibfield  {journal}
  {\bibinfo  {journal} {The Journal of Chemical Physics}\ }\textbf {\bibinfo
  {volume} {122}},\ \bibinfo {pages} {224103} (\bibinfo {year}
  {2005})}\BibitemShut {NoStop}%
\bibitem [{\citenamefont {Toher}\ \emph {et~al.}(2005)\citenamefont {Toher},
  \citenamefont {Filippetti}, \citenamefont {Sanvito},\ and\ \citenamefont
  {Burke}}]{toh}%
  \BibitemOpen
  \bibfield  {author} {\bibinfo {author} {\bibfnamefont {C.}~\bibnamefont
  {Toher}}, \bibinfo {author} {\bibfnamefont {A.}~\bibnamefont {Filippetti}},
  \bibinfo {author} {\bibfnamefont {S.}~\bibnamefont {Sanvito}}, \ and\
  \bibinfo {author} {\bibfnamefont {K.}~\bibnamefont {Burke}},\ }\bibfield
  {title} {\enquote {\bibinfo {title} {Self-interaction errors in
  density-functional calculations of electronic transport},}\ }\href {\doibase
  10.1103/PhysRevLett.95.146402} {\bibfield  {journal} {\bibinfo  {journal}
  {Physical Review Letters}\ }\textbf {\bibinfo {volume} {95}},\ \bibinfo
  {pages} {146402} (\bibinfo {year} {2005})}\BibitemShut {NoStop}%
\bibitem [{\citenamefont {Goedecker}\ and\ \citenamefont
  {Umrigar}(1997)}]{goed1}%
  \BibitemOpen
  \bibfield  {author} {\bibinfo {author} {\bibfnamefont {S.}~\bibnamefont
  {Goedecker}}\ and\ \bibinfo {author} {\bibfnamefont {C.}~\bibnamefont
  {Umrigar}},\ }\bibfield  {title} {\enquote {\bibinfo {title} {Critical
  assessment of the self-interaction-corrected--local-density-functional method
  and its algorithmic implementation},}\ }\href {\doibase
  10.1103/PhysRevA.55.1765} {\bibfield  {journal} {\bibinfo  {journal}
  {Physical Review A}\ }\textbf {\bibinfo {volume} {55}},\ \bibinfo {pages}
  {1765} (\bibinfo {year} {1997})}\BibitemShut {NoStop}%
\bibitem [{\citenamefont {Van~Leeuwen}\ and\ \citenamefont
  {Baerends}(1994)}]{van}%
  \BibitemOpen
  \bibfield  {author} {\bibinfo {author} {\bibfnamefont {R.}~\bibnamefont
  {Van~Leeuwen}}\ and\ \bibinfo {author} {\bibfnamefont {E.}~\bibnamefont
  {Baerends}},\ }\bibfield  {title} {\enquote {\bibinfo {title}
  {Exchange-correlation potential with correct asymptotic behavior},}\ }\href
  {\doibase 10.1103/PhysRevA.49.2421} {\bibfield  {journal} {\bibinfo
  {journal} {Physical Review A}\ }\textbf {\bibinfo {volume} {49}},\ \bibinfo
  {pages} {2421} (\bibinfo {year} {1994})}\BibitemShut {NoStop}%
\bibitem [{\citenamefont {Legrand}, \citenamefont {Suraud},\ and\ \citenamefont
  {Reinhard}(2002)}]{legr}%
  \BibitemOpen
  \bibfield  {author} {\bibinfo {author} {\bibfnamefont {C.}~\bibnamefont
  {Legrand}}, \bibinfo {author} {\bibfnamefont {E.}~\bibnamefont {Suraud}}, \
  and\ \bibinfo {author} {\bibfnamefont {P.}~\bibnamefont {Reinhard}},\
  }\bibfield  {title} {\enquote {\bibinfo {title} {Comparison of
  self-interaction-corrections for metal clusters},}\ }\href {\doibase
  10.1088/0953-4075/35/4/333} {\bibfield  {journal} {\bibinfo  {journal}
  {Journal of Physics B: Atomic, Molecular and Optical Physics}\ }\textbf
  {\bibinfo {volume} {35}},\ \bibinfo {pages} {1115} (\bibinfo {year}
  {2002})}\BibitemShut {NoStop}%
\bibitem [{\citenamefont {Tsuneda}\ and\ \citenamefont {Hirao}(2014)}]{tsu}%
  \BibitemOpen
  \bibfield  {author} {\bibinfo {author} {\bibfnamefont {T.}~\bibnamefont
  {Tsuneda}}\ and\ \bibinfo {author} {\bibfnamefont {K.}~\bibnamefont
  {Hirao}},\ }\bibfield  {title} {\enquote {\bibinfo {title} {Self-interaction
  corrections in density functional theory},}\ }\href {\doibase
  10.1063/1.4866996} {\bibfield  {journal} {\bibinfo  {journal} {The Journal of
  Chemical Physics}\ }\textbf {\bibinfo {volume} {140}},\ \bibinfo {pages}
  {18A513} (\bibinfo {year} {2014})}\BibitemShut {NoStop}%
\bibitem [{\citenamefont {Pederson}, \citenamefont {Ruzsinszky},\ and\
  \citenamefont {Perdew}(2014)}]{peders}%
  \BibitemOpen
  \bibfield  {author} {\bibinfo {author} {\bibfnamefont {M.~R.}\ \bibnamefont
  {Pederson}}, \bibinfo {author} {\bibfnamefont {A.}~\bibnamefont
  {Ruzsinszky}}, \ and\ \bibinfo {author} {\bibfnamefont {J.~P.}\ \bibnamefont
  {Perdew}},\ }\bibfield  {title} {\enquote {\bibinfo {title} {Communication:
  Self-interaction correction with unitary invariance in density functional
  theory},}\ }\href {\doibase 10.1063/1.4869581} {\bibfield  {journal}
  {\bibinfo  {journal} {The Journal of Chemical Physics}\ }\textbf {\bibinfo
  {volume} {140}},\ \bibinfo {pages} {121103} (\bibinfo {year}
  {2014})}\BibitemShut {NoStop}%
\bibitem [{\citenamefont {Gidopoulos}\ and\ \citenamefont
  {Lathiotakis}(2015{\natexlab{b}})}]{gid2015}%
  \BibitemOpen
  \bibfield  {author} {\bibinfo {author} {\bibfnamefont {N.}~\bibnamefont
  {Gidopoulos}}\ and\ \bibinfo {author} {\bibfnamefont {N.~N.}\ \bibnamefont
  {Lathiotakis}},\ }\bibfield  {title} {\enquote {\bibinfo {title} {Constrained
  local potentials for self-interaction correction},}\ }in\ \href {\doibase
  10.1016/bs.aamop.2015.06.003} {\emph {\bibinfo {booktitle} {Advances In
  Atomic, Molecular, and Optical Physics}}},\ Vol.~\bibinfo {volume} {64}\
  (\bibinfo  {publisher} {Elsevier},\ \bibinfo {year} {2015})\ pp.\ \bibinfo
  {pages} {129--142}\BibitemShut {NoStop}%
\bibitem [{\citenamefont {Clark}\ \emph {et~al.}(2017)\citenamefont {Clark},
  \citenamefont {Hollins}, \citenamefont {Refson},\ and\ \citenamefont
  {Gidopoulos}}]{clark}%
  \BibitemOpen
  \bibfield  {author} {\bibinfo {author} {\bibfnamefont {S.~J.}\ \bibnamefont
  {Clark}}, \bibinfo {author} {\bibfnamefont {T.~W.}\ \bibnamefont {Hollins}},
  \bibinfo {author} {\bibfnamefont {K.}~\bibnamefont {Refson}}, \ and\ \bibinfo
  {author} {\bibfnamefont {N.~I.}\ \bibnamefont {Gidopoulos}},\ }\bibfield
  {title} {\enquote {\bibinfo {title} {Self-interaction free local exchange
  potentials applied to metallic systems},}\ }\href {\doibase
  10.1088/1361-648X/aa7ba6} {\bibfield  {journal} {\bibinfo  {journal} {Journal
  of Physics: Condensed Matter}\ }\textbf {\bibinfo {volume} {29}},\ \bibinfo
  {pages} {374002} (\bibinfo {year} {2017})}\BibitemShut {NoStop}%
\bibitem [{\citenamefont {G{\"o}rling}(1999)}]{gorl}%
  \BibitemOpen
  \bibfield  {author} {\bibinfo {author} {\bibfnamefont {A.}~\bibnamefont
  {G{\"o}rling}},\ }\bibfield  {title} {\enquote {\bibinfo {title} {New ks
  method for molecules based on an exchange charge density generating the exact
  local ks exchange potential},}\ }\href {\doibase 10.1103/physrevlett.83.5459}
  {\bibfield  {journal} {\bibinfo  {journal} {Physical Review Letters}\
  }\textbf {\bibinfo {volume} {83}},\ \bibinfo {pages} {5459} (\bibinfo {year}
  {1999})}\BibitemShut {NoStop}%
\bibitem [{\citenamefont {Callow}, \citenamefont {Lathiotakis},\ and\
  \citenamefont {Gidopoulos}(2020)}]{inv1}%
  \BibitemOpen
  \bibfield  {author} {\bibinfo {author} {\bibfnamefont {T.~J.}\ \bibnamefont
  {Callow}}, \bibinfo {author} {\bibfnamefont {N.~N.}\ \bibnamefont
  {Lathiotakis}}, \ and\ \bibinfo {author} {\bibfnamefont {N.~I.}\ \bibnamefont
  {Gidopoulos}},\ }\bibfield  {title} {\enquote {\bibinfo {title}
  {Density-inversion method for the kohn–sham potential: Role of the
  screening density},}\ }\href {\doibase 10.1063/5.0005781} {\bibfield
  {journal} {\bibinfo  {journal} {The Journal of Chemical Physics}\ }\textbf
  {\bibinfo {volume} {152}},\ \bibinfo {pages} {164114} (\bibinfo {year}
  {2020})}\BibitemShut {NoStop}%
\bibitem [{\citenamefont {Bousiadi}, \citenamefont {Gidopoulos},\ and\
  \citenamefont {Lathiotakis}(2022)}]{inv2}%
  \BibitemOpen
  \bibfield  {author} {\bibinfo {author} {\bibfnamefont {S.}~\bibnamefont
  {Bousiadi}}, \bibinfo {author} {\bibfnamefont {N.}~\bibnamefont {Gidopoulos},
  \bibfnamefont {I}}, \ and\ \bibinfo {author} {\bibfnamefont {N.~N.}\
  \bibnamefont {Lathiotakis}},\ }\bibfield  {title} {\enquote {\bibinfo {title}
  {Density inversion method for local basis sets without potential auxiliary
  functions: inverting densities from rdmft},}\ }\href {\doibase
  10.1039/d2cp01866g} {\bibfield  {journal} {\bibinfo  {journal} {Physical
  Chemistry Chemical Physics}\ }\textbf {\bibinfo {volume} {24}},\ \bibinfo
  {pages} {19279--19286} (\bibinfo {year} {2022})}\BibitemShut {NoStop}%
\bibitem [{\citenamefont {Pitts}, \citenamefont {Lathiotakis},\ and\
  \citenamefont {Gidopoulos}(2021)}]{hybrid}%
  \BibitemOpen
  \bibfield  {author} {\bibinfo {author} {\bibfnamefont {T.~C.}\ \bibnamefont
  {Pitts}}, \bibinfo {author} {\bibfnamefont {N.~N.}\ \bibnamefont
  {Lathiotakis}}, \ and\ \bibinfo {author} {\bibfnamefont {N.}~\bibnamefont
  {Gidopoulos}},\ }\bibfield  {title} {\enquote {\bibinfo {title} {Generalized
  kohn--sham equations with accurate total energy and single-particle
  eigenvalue spectrum},}\ }\href {\doibase 10.1063/5.0071205} {\bibfield
  {journal} {\bibinfo  {journal} {The Journal of Chemical Physics}\ }\textbf
  {\bibinfo {volume} {155}},\ \bibinfo {pages} {224105} (\bibinfo {year}
  {2021})}\BibitemShut {NoStop}%
\bibitem [{\citenamefont {Lathiotakis}\ \emph
  {et~al.}(2014{\natexlab{a}})\citenamefont {Lathiotakis}, \citenamefont
  {Helbig}, \citenamefont {Rubio},\ and\ \citenamefont {Gidopoulos}}]{lathiot}%
  \BibitemOpen
  \bibfield  {author} {\bibinfo {author} {\bibfnamefont {N.~N.}\ \bibnamefont
  {Lathiotakis}}, \bibinfo {author} {\bibfnamefont {N.}~\bibnamefont {Helbig}},
  \bibinfo {author} {\bibfnamefont {A.}~\bibnamefont {Rubio}}, \ and\ \bibinfo
  {author} {\bibfnamefont {N.~I.}\ \bibnamefont {Gidopoulos}},\ }\bibfield
  {title} {\enquote {\bibinfo {title} {Local reduced-density-matrix-functional
  theory: Incorporating static correlation effects in kohn-sham equations},}\
  }\href {\doibase 10.1103/PhysRevA.90.032511} {\bibfield  {journal} {\bibinfo
  {journal} {Physical Review A}\ }\textbf {\bibinfo {volume} {90}},\ \bibinfo
  {pages} {032511} (\bibinfo {year} {2014}{\natexlab{a}})}\BibitemShut
  {NoStop}%
\bibitem [{\citenamefont {Lathiotakis}\ \emph
  {et~al.}(2014{\natexlab{b}})\citenamefont {Lathiotakis}, \citenamefont
  {Helbig}, \citenamefont {Rubio},\ and\ \citenamefont
  {Gidopoulos}}]{lathiot2}%
  \BibitemOpen
  \bibfield  {author} {\bibinfo {author} {\bibfnamefont {N.~N.}\ \bibnamefont
  {Lathiotakis}}, \bibinfo {author} {\bibfnamefont {N.}~\bibnamefont {Helbig}},
  \bibinfo {author} {\bibfnamefont {A.}~\bibnamefont {Rubio}}, \ and\ \bibinfo
  {author} {\bibfnamefont {N.~I.}\ \bibnamefont {Gidopoulos}},\ }\bibfield
  {title} {\enquote {\bibinfo {title} {Quasi-particle energy spectra in local
  reduced density matrix functional theory},}\ }\href {\doibase
  10.1063/1.4899072} {\bibfield  {journal} {\bibinfo  {journal} {The Journal of
  Chemical Physics}\ }\textbf {\bibinfo {volume} {141}},\ \bibinfo {pages}
  {164120} (\bibinfo {year} {2014}{\natexlab{b}})}\BibitemShut {NoStop}%
\bibitem [{\citenamefont {Gilbert}(1975)}]{gilb}%
  \BibitemOpen
  \bibfield  {author} {\bibinfo {author} {\bibfnamefont {T.~L.}\ \bibnamefont
  {Gilbert}},\ }\bibfield  {title} {\enquote {\bibinfo {title} {Hohenberg-kohn
  theorem for nonlocal external potentials},}\ }\href {\doibase
  10.1103/PhysRevB.12.2111} {\bibfield  {journal} {\bibinfo  {journal}
  {Physical Review B}\ }\textbf {\bibinfo {volume} {12}},\ \bibinfo {pages}
  {2111} (\bibinfo {year} {1975})}\BibitemShut {NoStop}%
\bibitem [{\citenamefont {Zumbach}\ and\ \citenamefont {Maschke}(1985)}]{zumb}%
  \BibitemOpen
  \bibfield  {author} {\bibinfo {author} {\bibfnamefont {G.}~\bibnamefont
  {Zumbach}}\ and\ \bibinfo {author} {\bibfnamefont {K.}~\bibnamefont
  {Maschke}},\ }\bibfield  {title} {\enquote {\bibinfo {title} {Density-matrix
  functional theory for the n-particle ground state},}\ }\href {\doibase
  10.1063/1.448595} {\bibfield  {journal} {\bibinfo  {journal} {The Journal of
  Chemical Physics}\ }\textbf {\bibinfo {volume} {82}},\ \bibinfo {pages}
  {5604--5607} (\bibinfo {year} {1985})}\BibitemShut {NoStop}%
\bibitem [{\citenamefont {Levy}(1979)}]{levy}%
  \BibitemOpen
  \bibfield  {author} {\bibinfo {author} {\bibfnamefont {M.}~\bibnamefont
  {Levy}},\ }\bibfield  {title} {\enquote {\bibinfo {title} {Universal
  variational functionals of electron densities, first-order density matrices,
  and natural spin-orbitals and solution of the v-representability problem},}\
  }\href {\doibase 10.1073/pnas.76.12.6062} {\bibfield  {journal} {\bibinfo
  {journal} {Proceedings of the National Academy of Sciences}\ }\textbf
  {\bibinfo {volume} {76}},\ \bibinfo {pages} {6062--6065} (\bibinfo {year}
  {1979})}\BibitemShut {NoStop}%
\bibitem [{\citenamefont {Goedecker}\ and\ \citenamefont
  {Umrigar}(1998)}]{goed2}%
  \BibitemOpen
  \bibfield  {author} {\bibinfo {author} {\bibfnamefont {S.}~\bibnamefont
  {Goedecker}}\ and\ \bibinfo {author} {\bibfnamefont {C.}~\bibnamefont
  {Umrigar}},\ }\bibfield  {title} {\enquote {\bibinfo {title} {Natural orbital
  functional for the many-electron problem},}\ }\href {\doibase
  10.1103/PhysRevLett.81.866} {\bibfield  {journal} {\bibinfo  {journal}
  {Physical Review Letters}\ }\textbf {\bibinfo {volume} {81}},\ \bibinfo
  {pages} {866} (\bibinfo {year} {1998})}\BibitemShut {NoStop}%
\bibitem [{\citenamefont {Buijse}\ and\ \citenamefont {Baerends}(2002)}]{bu}%
  \BibitemOpen
  \bibfield  {author} {\bibinfo {author} {\bibfnamefont {M.}~\bibnamefont
  {Buijse}}\ and\ \bibinfo {author} {\bibfnamefont {E.}~\bibnamefont
  {Baerends}},\ }\bibfield  {title} {\enquote {\bibinfo {title} {An approximate
  exchange-correlation hole density as a functional of the natural orbitals},}\
  }\href {\doibase 10.1080/00268970110070243} {\bibfield  {journal} {\bibinfo
  {journal} {Molecular Physics}\ }\textbf {\bibinfo {volume} {100}},\ \bibinfo
  {pages} {401--421} (\bibinfo {year} {2002})}\BibitemShut {NoStop}%
\bibitem [{\citenamefont {Gritsenko}, \citenamefont {Pernal},\ and\
  \citenamefont {Baerends}(2005)}]{BB3}%
  \BibitemOpen
  \bibfield  {author} {\bibinfo {author} {\bibfnamefont {O.}~\bibnamefont
  {Gritsenko}}, \bibinfo {author} {\bibfnamefont {K.}~\bibnamefont {Pernal}}, \
  and\ \bibinfo {author} {\bibfnamefont {E.~J.}\ \bibnamefont {Baerends}},\
  }\bibfield  {title} {\enquote {\bibinfo {title} {An improved density matrix
  functional by physically motivated repulsive corrections},}\ }\href {\doibase
  10.1063/1.1906203} {\bibfield  {journal} {\bibinfo  {journal} {The Journal of
  Chemical Physics}\ }\textbf {\bibinfo {volume} {122}},\ \bibinfo {pages}
  {204102} (\bibinfo {year} {2005})}\BibitemShut {NoStop}%
\bibitem [{\citenamefont {Lathiotakis}\ \emph {et~al.}(2009)\citenamefont
  {Lathiotakis}, \citenamefont {Sharma}, \citenamefont {Dewhurst},
  \citenamefont {Eich}, \citenamefont {Marques},\ and\ \citenamefont
  {Gross}}]{lathiotakis2009density}%
  \BibitemOpen
  \bibfield  {author} {\bibinfo {author} {\bibfnamefont {N.}~\bibnamefont
  {Lathiotakis}}, \bibinfo {author} {\bibfnamefont {S.}~\bibnamefont {Sharma}},
  \bibinfo {author} {\bibfnamefont {J.}~\bibnamefont {Dewhurst}}, \bibinfo
  {author} {\bibfnamefont {F.~G.}\ \bibnamefont {Eich}}, \bibinfo {author}
  {\bibfnamefont {M.}~\bibnamefont {Marques}}, \ and\ \bibinfo {author}
  {\bibfnamefont {E.}~\bibnamefont {Gross}},\ }\bibfield  {title} {\enquote
  {\bibinfo {title} {Density-matrix-power functional: Performance for finite
  systems and the homogeneous electron gas},}\ }\href {\doibase
  10.1103/PhysRevA.79.040501} {\bibfield  {journal} {\bibinfo  {journal}
  {Physical Review A}\ }\textbf {\bibinfo {volume} {79}},\ \bibinfo {pages}
  {040501} (\bibinfo {year} {2009})}\BibitemShut {NoStop}%
\bibitem [{\citenamefont {Marques}\ and\ \citenamefont
  {Lathiotakis}(2008)}]{ML}%
  \BibitemOpen
  \bibfield  {author} {\bibinfo {author} {\bibfnamefont {M.~A.}\ \bibnamefont
  {Marques}}\ and\ \bibinfo {author} {\bibfnamefont {N.}~\bibnamefont
  {Lathiotakis}},\ }\bibfield  {title} {\enquote {\bibinfo {title} {Empirical
  functionals for reduced-density-matrix-functional theory},}\ }\href {\doibase
  10.1103/PhysRevA.77.032509} {\bibfield  {journal} {\bibinfo  {journal}
  {Physical Review A}\ }\textbf {\bibinfo {volume} {77}},\ \bibinfo {pages}
  {032509} (\bibinfo {year} {2008})}\BibitemShut {NoStop}%
\bibitem [{\citenamefont {Piris}\ and\ \citenamefont {Ugalde}(2009)}]{piris}%
  \BibitemOpen
  \bibfield  {author} {\bibinfo {author} {\bibfnamefont {M.}~\bibnamefont
  {Piris}}\ and\ \bibinfo {author} {\bibfnamefont {J.~M.}\ \bibnamefont
  {Ugalde}},\ }\bibfield  {title} {\enquote {\bibinfo {title} {Iterative
  diagonalization for orbital optimization in natural orbital functional
  theory},}\ }\href {\doibase 10.1002/jcc.21225} {\bibfield  {journal}
  {\bibinfo  {journal} {Journal of Computational Chemistry}\ }\textbf {\bibinfo
  {volume} {30}},\ \bibinfo {pages} {2078--2086} (\bibinfo {year}
  {2009})}\BibitemShut {NoStop}%
\bibitem [{\citenamefont {Pernal}, \citenamefont {Gritsenko},\ and\
  \citenamefont {Baerends}(2007)}]{ext1}%
  \BibitemOpen
  \bibfield  {author} {\bibinfo {author} {\bibfnamefont {K.}~\bibnamefont
  {Pernal}}, \bibinfo {author} {\bibfnamefont {O.}~\bibnamefont {Gritsenko}}, \
  and\ \bibinfo {author} {\bibfnamefont {E.~J.}\ \bibnamefont {Baerends}},\
  }\bibfield  {title} {\enquote {\bibinfo {title} {Time-dependent
  density-matrix-functional theory},}\ }\href {\doibase
  10.1103/PhysRevA.75.012506} {\bibfield  {journal} {\bibinfo  {journal} {Phys.
  Rev. A}\ }\textbf {\bibinfo {volume} {75}},\ \bibinfo {pages} {012506}
  (\bibinfo {year} {2007})}\BibitemShut {NoStop}%
\bibitem [{\citenamefont {Baldsiefen}, \citenamefont {Cangi},\ and\
  \citenamefont {Gross}(2015)}]{ext2}%
  \BibitemOpen
  \bibfield  {author} {\bibinfo {author} {\bibfnamefont {T.}~\bibnamefont
  {Baldsiefen}}, \bibinfo {author} {\bibfnamefont {A.}~\bibnamefont {Cangi}}, \
  and\ \bibinfo {author} {\bibfnamefont {E.~K.~U.}\ \bibnamefont {Gross}},\
  }\bibfield  {title} {\enquote {\bibinfo {title}
  {Reduced-density-matrix-functional theory at finite temperature: Theoretical
  foundations},}\ }\href {\doibase 10.1103/PhysRevA.92.052514} {\bibfield
  {journal} {\bibinfo  {journal} {Phys. Rev. A}\ }\textbf {\bibinfo {volume}
  {92}},\ \bibinfo {pages} {052514} (\bibinfo {year} {2015})}\BibitemShut
  {NoStop}%
\bibitem [{\citenamefont {Liebert}\ \emph {et~al.}(2022)\citenamefont
  {Liebert}, \citenamefont {Castillo}, \citenamefont {Labbé},\ and\
  \citenamefont {Schilling}}]{ext3}%
  \BibitemOpen
  \bibfield  {author} {\bibinfo {author} {\bibfnamefont {J.}~\bibnamefont
  {Liebert}}, \bibinfo {author} {\bibfnamefont {F.}~\bibnamefont {Castillo}},
  \bibinfo {author} {\bibfnamefont {J.-P.}\ \bibnamefont {Labbé}}, \ and\
  \bibinfo {author} {\bibfnamefont {C.}~\bibnamefont {Schilling}},\ }\bibfield
  {title} {\enquote {\bibinfo {title} {Foundation of one-particle reduced
  density matrix functional theory for excited states},}\ }\href {\doibase
  10.1021/acs.jctc.1c00561} {\bibfield  {journal} {\bibinfo  {journal} {Journal
  of Chemical Theory and Computation}\ }\textbf {\bibinfo {volume} {18}},\
  \bibinfo {pages} {124--140} (\bibinfo {year} {2022})}\BibitemShut {NoStop}%
\bibitem [{\citenamefont {Coleman}(1972)}]{coleman}%
  \BibitemOpen
  \bibfield  {author} {\bibinfo {author} {\bibfnamefont {A.}~\bibnamefont
  {Coleman}},\ }\bibfield  {title} {\enquote {\bibinfo {title} {Necessary
  conditions for n-representability of reduced density matrices},}\ }\href
  {\doibase 10.1063/1.1665956} {\bibfield  {journal} {\bibinfo  {journal}
  {Journal of Mathematical Physics}\ }\textbf {\bibinfo {volume} {13}},\
  \bibinfo {pages} {214--222} (\bibinfo {year} {1972})}\BibitemShut {NoStop}%
\bibitem [{\citenamefont {Cs{\'a}nyi}, \citenamefont {Goedecker},\ and\
  \citenamefont {Arias}(2002)}]{fun4}%
  \BibitemOpen
  \bibfield  {author} {\bibinfo {author} {\bibfnamefont {G.}~\bibnamefont
  {Cs{\'a}nyi}}, \bibinfo {author} {\bibfnamefont {S.}~\bibnamefont
  {Goedecker}}, \ and\ \bibinfo {author} {\bibfnamefont {T.}~\bibnamefont
  {Arias}},\ }\bibfield  {title} {\enquote {\bibinfo {title} {Improved
  tensor-product expansions for the two-particle density matrix},}\ }\href
  {\doibase 10.1103/PhysRevA.65.032510} {\bibfield  {journal} {\bibinfo
  {journal} {Physical Review A}\ }\textbf {\bibinfo {volume} {65}},\ \bibinfo
  {pages} {032510} (\bibinfo {year} {2002})}\BibitemShut {NoStop}%
\bibitem [{\citenamefont {Kollmar}(2004)}]{fun5}%
  \BibitemOpen
  \bibfield  {author} {\bibinfo {author} {\bibfnamefont {C.}~\bibnamefont
  {Kollmar}},\ }\bibfield  {title} {\enquote {\bibinfo {title} {The
  “jk-only” approximation in density matrix functional and wave function
  theory},}\ }\href {\doibase 10.1063/1.1819319} {\bibfield  {journal}
  {\bibinfo  {journal} {The Journal of Chemical Physics}\ }\textbf {\bibinfo
  {volume} {121}},\ \bibinfo {pages} {11581--11586} (\bibinfo {year}
  {2004})}\BibitemShut {NoStop}%
\bibitem [{\citenamefont {Piris}(2006)}]{fun6}%
  \BibitemOpen
  \bibfield  {author} {\bibinfo {author} {\bibfnamefont {M.}~\bibnamefont
  {Piris}},\ }\bibfield  {title} {\enquote {\bibinfo {title} {A new approach
  for the two-electron cumulant in natural orbital functional theory},}\ }\href
  {\doibase 10.1002/qua.20858} {\bibfield  {journal} {\bibinfo  {journal}
  {International Journal of Quantum Chemistry}\ }\textbf {\bibinfo {volume}
  {106}},\ \bibinfo {pages} {1093--1104} (\bibinfo {year} {2006})}\BibitemShut
  {NoStop}%
\bibitem [{\citenamefont {Sharma}\ \emph {et~al.}(2008)\citenamefont {Sharma},
  \citenamefont {Dewhurst}, \citenamefont {Lathiotakis},\ and\ \citenamefont
  {Gross}}]{pow}%
  \BibitemOpen
  \bibfield  {author} {\bibinfo {author} {\bibfnamefont {S.}~\bibnamefont
  {Sharma}}, \bibinfo {author} {\bibfnamefont {J.~K.}\ \bibnamefont
  {Dewhurst}}, \bibinfo {author} {\bibfnamefont {N.~N.}\ \bibnamefont
  {Lathiotakis}}, \ and\ \bibinfo {author} {\bibfnamefont {E.~K.}\ \bibnamefont
  {Gross}},\ }\bibfield  {title} {\enquote {\bibinfo {title} {Reduced density
  matrix functional for many-electron systems},}\ }\href {\doibase
  10.1103/PhysRevB.78.201103} {\bibfield  {journal} {\bibinfo  {journal}
  {Physical Review B}\ }\textbf {\bibinfo {volume} {78}},\ \bibinfo {pages}
  {201103} (\bibinfo {year} {2008})}\BibitemShut {NoStop}%
\bibitem [{\citenamefont {M{\"u}ller}(1984)}]{BB0}%
  \BibitemOpen
  \bibfield  {author} {\bibinfo {author} {\bibfnamefont {A.}~\bibnamefont
  {M{\"u}ller}},\ }\bibfield  {title} {\enquote {\bibinfo {title} {Explicit
  approximate relation between reduced two-and one-particle density
  matrices},}\ }\href {\doibase 10.1016/0375-9601(84)91034-X} {\bibfield
  {journal} {\bibinfo  {journal} {Physics Letters A}\ }\textbf {\bibinfo
  {volume} {105}},\ \bibinfo {pages} {446--452} (\bibinfo {year}
  {1984})}\BibitemShut {NoStop}%
\bibitem [{\citenamefont {Piris}\ \emph {et~al.}(2011)\citenamefont {Piris},
  \citenamefont {Lopez}, \citenamefont {Ruip{\'e}rez}, \citenamefont
  {Matxain},\ and\ \citenamefont {Ugalde}}]{pn5}%
  \BibitemOpen
  \bibfield  {author} {\bibinfo {author} {\bibfnamefont {M.}~\bibnamefont
  {Piris}}, \bibinfo {author} {\bibfnamefont {X.}~\bibnamefont {Lopez}},
  \bibinfo {author} {\bibfnamefont {F.}~\bibnamefont {Ruip{\'e}rez}}, \bibinfo
  {author} {\bibfnamefont {J.}~\bibnamefont {Matxain}}, \ and\ \bibinfo
  {author} {\bibfnamefont {J.}~\bibnamefont {Ugalde}},\ }\bibfield  {title}
  {\enquote {\bibinfo {title} {A natural orbital functional for
  multiconfigurational states},}\ }\href {\doibase 10.1063/1.3582792}
  {\bibfield  {journal} {\bibinfo  {journal} {The Journal of Chemical Physics}\
  }\textbf {\bibinfo {volume} {134}},\ \bibinfo {pages} {164102} (\bibinfo
  {year} {2011})}\BibitemShut {NoStop}%
\bibitem [{\citenamefont {Piris}(2021)}]{PhysRevLett.127.233001}%
  \BibitemOpen
  \bibfield  {author} {\bibinfo {author} {\bibfnamefont {M.}~\bibnamefont
  {Piris}},\ }\bibfield  {title} {\enquote {\bibinfo {title} {Global natural
  orbital functional: Towards the complete description of the electron
  correlation},}\ }\href {\doibase 10.1103/PhysRevLett.127.233001} {\bibfield
  {journal} {\bibinfo  {journal} {Physical Review Letters}\ }\textbf {\bibinfo
  {volume} {127}},\ \bibinfo {pages} {233001} (\bibinfo {year}
  {2021})}\BibitemShut {NoStop}%
\bibitem [{\citenamefont {Theophilou}\ \emph {et~al.}(2015)\citenamefont
  {Theophilou}, \citenamefont {Lathiotakis}, \citenamefont {Gidopoulos},
  \citenamefont {Rubio},\ and\ \citenamefont
  {Helbig}}]{theophilou2015orbitals}%
  \BibitemOpen
  \bibfield  {author} {\bibinfo {author} {\bibfnamefont {I.}~\bibnamefont
  {Theophilou}}, \bibinfo {author} {\bibfnamefont {N.~N.}\ \bibnamefont
  {Lathiotakis}}, \bibinfo {author} {\bibfnamefont {N.~I.}\ \bibnamefont
  {Gidopoulos}}, \bibinfo {author} {\bibfnamefont {A.}~\bibnamefont {Rubio}}, \
  and\ \bibinfo {author} {\bibfnamefont {N.}~\bibnamefont {Helbig}},\
  }\bibfield  {title} {\enquote {\bibinfo {title} {Orbitals from local rdmft:
  Are they kohn-sham or natural orbitals?}}\ }\href {\doibase
  10.1063/1.4927784} {\bibfield  {journal} {\bibinfo  {journal} {The Journal of
  Chemical Physics}\ }\textbf {\bibinfo {volume} {143}},\ \bibinfo {pages}
  {054106} (\bibinfo {year} {2015})}\BibitemShut {NoStop}%
\bibitem [{nis()}]{nist}%
  \BibitemOpen
  \href@noop {} {}\bibinfo {howpublished} {NIST Webbook,
  \url{http://webbook.nist.gov/chemistry/}}\BibitemShut {NoStop}%
\bibitem [{gsl()}]{gsl}%
  \BibitemOpen
  \href {http://www.gnu.org/software/gsl/} {}\bibinfo {howpublished} {M.
  Galassi et al, GNU Scientific Library Reference Manual (3rd Ed.), ISBN
  0954612078.}\BibitemShut {Stop}%
\bibitem [{hip()}]{hip}%
  \BibitemOpen
  \href@noop {} {}\bibinfo {howpublished} {HIPPO computer program,
  \url{lathiot@eie.gr}}\BibitemShut {NoStop}%
\bibitem [{\citenamefont {Hamel}\ \emph {et~al.}(2002)\citenamefont {Hamel},
  \citenamefont {Duffy}, \citenamefont {Casida},\ and\ \citenamefont
  {Salahub}}]{HAMEL2002345}%
  \BibitemOpen
  \bibfield  {author} {\bibinfo {author} {\bibfnamefont {S.}~\bibnamefont
  {Hamel}}, \bibinfo {author} {\bibfnamefont {P.}~\bibnamefont {Duffy}},
  \bibinfo {author} {\bibfnamefont {M.~E.}\ \bibnamefont {Casida}}, \ and\
  \bibinfo {author} {\bibfnamefont {D.~R.}\ \bibnamefont {Salahub}},\
  }\bibfield  {title} {\enquote {\bibinfo {title} {Kohn–sham orbitals and
  orbital energies: fictitious constructs but good approximations all the
  same},}\ }\href {\doibase https://doi.org/10.1016/S0368-2048(02)00032-4}
  {\bibfield  {journal} {\bibinfo  {journal} {Journal of Electron Spectroscopy
  and Related Phenomena}\ }\textbf {\bibinfo {volume} {123}},\ \bibinfo {pages}
  {345--363} (\bibinfo {year} {2002})}\BibitemShut {NoStop}%
\bibitem [{\citenamefont {Chong}, \citenamefont {Gritsenko},\ and\
  \citenamefont {Baerends}(2002)}]{doi:10.1063/1.1430255}%
  \BibitemOpen
  \bibfield  {author} {\bibinfo {author} {\bibfnamefont {D.~P.}\ \bibnamefont
  {Chong}}, \bibinfo {author} {\bibfnamefont {O.~V.}\ \bibnamefont
  {Gritsenko}}, \ and\ \bibinfo {author} {\bibfnamefont {E.~J.}\ \bibnamefont
  {Baerends}},\ }\bibfield  {title} {\enquote {\bibinfo {title} {Interpretation
  of the kohn–sham orbital energies as approximate vertical ionization
  potentials},}\ }\href {\doibase 10.1063/1.1430255} {\bibfield  {journal}
  {\bibinfo  {journal} {The Journal of Chemical Physics}\ }\textbf {\bibinfo
  {volume} {116}},\ \bibinfo {pages} {1760--1772} (\bibinfo {year}
  {2002})}\BibitemShut {NoStop}%
\bibitem [{\citenamefont {van Meer}, \citenamefont {Gritsenko},\ and\
  \citenamefont {Baerends}(2014)}]{WOS:000343196300027}%
  \BibitemOpen
  \bibfield  {author} {\bibinfo {author} {\bibfnamefont {R.}~\bibnamefont {van
  Meer}}, \bibinfo {author} {\bibfnamefont {O.~V.}\ \bibnamefont {Gritsenko}},
  \ and\ \bibinfo {author} {\bibfnamefont {E.~J.}\ \bibnamefont {Baerends}},\
  }\bibfield  {title} {\enquote {\bibinfo {title} {Physical meaning of virtual
  kohn-sham orbitals and orbital energies: An ideal basis for the description
  of molecular excitations},}\ }\href {\doibase 10.1021/ct500727c} {\bibfield
  {journal} {\bibinfo  {journal} {Journal of Chemical Theory and Computation}\
  }\textbf {\bibinfo {volume} {10}},\ \bibinfo {pages} {4432--4441} (\bibinfo
  {year} {2014})}\BibitemShut {NoStop}%
\bibitem [{\citenamefont {Verma}\ and\ \citenamefont
  {Bartlett}(2012)}]{doi:10.1063/1.3678180}%
  \BibitemOpen
  \bibfield  {author} {\bibinfo {author} {\bibfnamefont {P.}~\bibnamefont
  {Verma}}\ and\ \bibinfo {author} {\bibfnamefont {R.~J.}\ \bibnamefont
  {Bartlett}},\ }\bibfield  {title} {\enquote {\bibinfo {title} {Increasing the
  applicability of density functional theory. ii. correlation potentials from
  the random phase approximation and beyond},}\ }\href {\doibase
  10.1063/1.3678180} {\bibfield  {journal} {\bibinfo  {journal} {The Journal of
  Chemical Physics}\ }\textbf {\bibinfo {volume} {136}},\ \bibinfo {pages}
  {044105} (\bibinfo {year} {2012})}\BibitemShut {NoStop}%
\bibitem [{\citenamefont {Bartlett}(2019)}]{doi:10.1063/1.5116338}%
  \BibitemOpen
  \bibfield  {author} {\bibinfo {author} {\bibfnamefont {R.~J.}\ \bibnamefont
  {Bartlett}},\ }\bibfield  {title} {\enquote {\bibinfo {title} {Adventures in
  dft by a wavefunction theorist},}\ }\href {\doibase 10.1063/1.5116338}
  {\bibfield  {journal} {\bibinfo  {journal} {The Journal of Chemical Physics}\
  }\textbf {\bibinfo {volume} {151}},\ \bibinfo {pages} {160901} (\bibinfo
  {year} {2019})}\BibitemShut {NoStop}%
\bibitem [{\citenamefont {Ranasinghe}\ \emph {et~al.}(2019)\citenamefont
  {Ranasinghe}, \citenamefont {Margraf}, \citenamefont {Perera},\ and\
  \citenamefont {Bartlett}}]{doi:10.1063/1.5084728}%
  \BibitemOpen
  \bibfield  {author} {\bibinfo {author} {\bibfnamefont {D.~S.}\ \bibnamefont
  {Ranasinghe}}, \bibinfo {author} {\bibfnamefont {J.~T.}\ \bibnamefont
  {Margraf}}, \bibinfo {author} {\bibfnamefont {A.}~\bibnamefont {Perera}}, \
  and\ \bibinfo {author} {\bibfnamefont {R.~J.}\ \bibnamefont {Bartlett}},\
  }\bibfield  {title} {\enquote {\bibinfo {title} {Vertical valence ionization
  potential benchmarks from equation-of-motion coupled cluster theory and qtp
  functionals},}\ }\href {\doibase 10.1063/1.5084728} {\bibfield  {journal}
  {\bibinfo  {journal} {The Journal of Chemical Physics}\ }\textbf {\bibinfo
  {volume} {150}},\ \bibinfo {pages} {074108} (\bibinfo {year}
  {2019})}\BibitemShut {NoStop}%
\end{thebibliography}%

\end{document}